\documentclass[acmsmall,screen]{acmart}

\AtBeginDocument{%
  }

\usepackage{natbib}

\usepackage{amsmath}
\usepackage{algorithmic}
\usepackage{adjustbox}
\usepackage{graphicx}
\usepackage{textcomp}
\usepackage{xcolor}
\usepackage{enumitem}
\usepackage{xspace}
\usepackage{hyperref}
\usepackage{multirow}
\usepackage{framed}
\usepackage[most,skins]{tcolorbox}
\usepackage{diagbox}
\usepackage{pifont}
\usepackage{subcaption}
\usepackage{orcidlink}

\usepackage{colortbl}
\usepackage{url} 
\usepackage{listings}
\usepackage{wrapfig}

\setcopyright{acmlicensed}
\copyrightyear{2026}

\acmYear{2026}
\acmVolume{1}
\acmArticle{1}

\acmISBN{978-1-4503-XXXX-X/2018/06}

\setcopyright{none}
\renewcommand\footnotetextcopyrightpermission[1]{} 

\begin{document}

\newcommand{\code}[1]{\textsc{#1}}
\newcommand{\todo}[1]{{\color{red} \bf \{TODO: {#1}\}}}

\newcommand{\wb}[1]{{\color{brown}[Bo: #1]}}

\def\modify#1#2#3{{\small{\sf{#1}}} {\color{red}{#2}}
{{\color{red}\mbox{$\Rightarrow$}}} {\color{blue}{#3}}}
\renewcommand{\modify}[3]{{#3}}

\newcommand{\mycomment}[2]{{\color{magenta}{\sf{#1}}} {\color{blue}{#2}}}
\renewcommand{\mycomment}[2]{{#2}}

\newcommand{\bocomment}[1]{\mycomment{}{#1}}
\newcommand{\bomodify}[2]{\modify{}{#1}{#2}}
\newcommand{\bomodifyno}[2]{{#1}}

\newcommand{\ourtoolraw}{LiAgent}
\newcommand{\ourtool}{\textsc{\ourtoolraw{}}\xspace}

\newcommand{\listref}[1]{List~\ref{lst:#1}}
\newcommand{\tabref}[1]{Table~\ref{tab:#1}}
\newcommand{\tablabel}[1]{\label{tab:#1}}
\newcommand{\figref}[1]{Figure~\ref{fig:#1}}
\newcommand{\figlabel}[1]{\label{fig:#1}}

\newcommand{\smalltitle}[1]{{\vspace{0.1cm} \noindent \bf  {#1}.\ }}

\newcommand{\smalltitlecolon}[1]{{\smallskip \noindent \bf  {#1}:\ }}

\newcommand{\finding}[1]{ \begin{tcolorbox}[tile, size=fbox,left=3mm, right=1mm, boxrule=0pt, top=1mm, bottom=1mm,
    borderline west={2mm}{0pt}{blue!70!black}, colback=blue!3!white, 
    sharp corners=south]
\textbf{Finding \refstepcounter{num}\thenum}: #1
\end{tcolorbox}}

\newcounter{num}

\title{Hidden Licensing Risks in the LLMware Ecosystem}

\author{Bo Wang}
\orcid{0000-0001-7944-9182}
\affiliation{
  \institution{Beijing Jiaotong University}
  \city{Beijing}
  \country{China}
}
\email{wangbo\_cs@bjtu.edu.cn}

\author{Yueyang Chen}
\orcid{0009-0004-0044-5090}
\affiliation{
  \institution{Beijing Jiaotong University}
  \city{Beijing}
  \country{China}
}
\email{24125217@bjtu.edu.cn}

\author{Jieke Shi}
\orcid{0000-0002-0799-5018}
\affiliation{
  \institution{Singapore Management University}
  \country{Singapore}
}
\email{jiekeshi@smu.edu.sg}

\author{Minghui Li}
\orcid{0009-0004-0421-3561}
\affiliation{
  \institution{Beijing Jiaotong University}
  \city{Beijing}
  \country{China}
}
\email{24125240@bjtu.edu.cn}

\author{Yunbo Lyu}
\orcid{0009-0004-2522-7348}
\affiliation{
  \institution{Singapore Management University}
  \country{Singapore}
}
\email{yunbolyu@smu.edu.sg}

\author{Yinan Wu}
\orcid{0009-0002-1005-2560}
\affiliation{
  \institution{North Carolina State University}
  \country{USA}
}
\email{ywu92@ncsu.edu}

\author{Youfang Lin}
\orcid{0000-0002-5143-3645}
\affiliation{
  \institution{Beijing Jiaotong University}
  \city{Beijing}
  \country{China}
}
\email{yflin@bjtu.edu.cn}

\author{Zhou Yang}
\orcid{0000-0001-5938-1918}
\affiliation{
  \institution{University of Alberta}
  \city{Alberta}
  \country{Canada}
}
\email{zy25@ualberta.ca}

\renewcommand{\shortauthors}{Wang et al.}

\begin{abstract}
Large Language Models (LLMs) have been increasingly integrated into software systems, giving rise to a new class of software referred to as \textit{LLMware}. In addition to traditional software components composed solely of source code, LLMware also embeds or interacts with LLMs that depend on other models and datasets, forming complex supply chains involving open-source software (OSS) libraries, LLMs, and datasets. However, the licensing issues arising from these intertwined dependencies remain largely unexplored.

Leveraging GitHub [\includegraphics[height=0.8em]{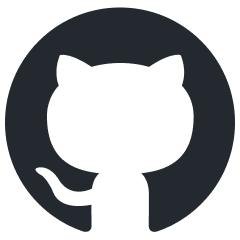}] and Hugging Face [\includegraphics[height=1em]{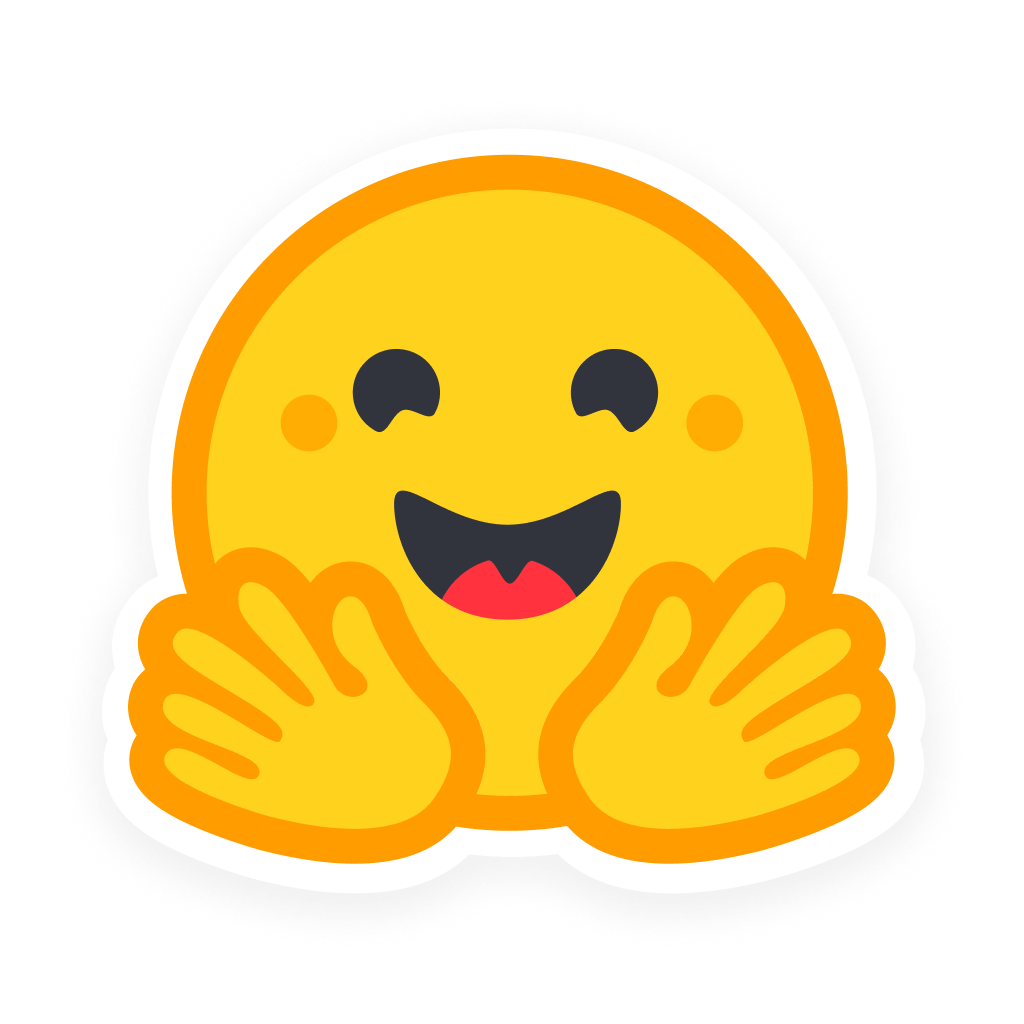}], two premier hubs for code and models, we curate a large-scale dataset capturing the supply chains of LLMware. Our dataset comprises 12,180 OSS repositories from GitHub, 3,988 LLMs, and 708 datasets from Hugging Face.
We analyze license distributions in the LLMware ecosystem and find that licensing practices differ markedly from those in traditional OSS communities. We further examine license-related issues and identify license selection and maintenance as the primary pain points, with 84\% of cases involving discussions about adding appropriate licenses or resolving conflicts in existing ones.
We then study license incompatibility in LLMware and evaluate the state-of-the-art approaches, finding that they perform poorly in this setting and achieve only 58\% and 76\% F1 scores, respectively. These results motivate us to propose \ourtool, which explores the potential of LLM-based agents for ecosystem-level license compatibility analysis, achieves an F1 score of 87\%, and improves performance by 14 percentage points over prior approaches. We submit 60 license incompatibility issues detected by \ourtool, of which 11 have been confirmed by developers.
Two LLMs with license conflicts have more than 107 million and 5 million downloads on Hugging Face, respectively, suggesting the issues may impact a large number of downstream applications.
We conclude by discussing implications and providing recommendations to support the healthy growth of the LLMware ecosystem.
\end{abstract}



\maketitle

\section{Introduction}
\label{sec:intro}

Large Language Models (LLMs) have achieved remarkable capabilities across a wide spectrum of tasks, from natural language dialogue and content creation~\cite{10.1145/3771090,10.1145/3641289,zhao2025surveylargelanguagemodels} to complex program generation and analysis~\cite{10449667,10.1145/3695988,chen2025deep,wang2024software}, driving the emergence of a new class of applications that incorporate LLMs or invoke them programmatically as core components. We refer to this emerging category as \textit{LLMware}. Prominent examples include conversational applications like ChatGPT~\cite{chatgpt} and AI-assisted development tools like GitHub Copilot~\cite{copilot}, which streamline workflows and improve productivity. These applications are fueling a rapidly expanding market projected to reach \$24.9B by 2031~\cite{mordorintelligence}, illustrating the growing role of LLMware in modern software ecosystems.

Software development is not an isolated endeavor but is shaped by a dense network of dependencies. Developers commonly build upon libraries and components from open-source software (OSS), allowing them to avoid reinventing the wheel and focus on their unique workflows and features~\cite{li2025open,10.1145/130844.130856,10.1145/3762635,10.1145/3696630.3728525}. Collectively, these dependencies form the software supply chain, i.e., the set of upstream components and artifacts on which a software depends~\cite{10.1145/3714464,10.1145/3708531}. However, unlike traditional software, whose supply chains primarily consist of OSS artifacts in the form of source code, LLMware introduces additional dependencies on the LLMs themselves, which may in turn rely on other models and datasets. LLMware’s supply chain thus becomes more complex~\cite{li2025open,10.1145/3708531,liu2025empiricalstudyvulnerablepackage}, and gives rise to a new licensing landscape across heterogeneous artifacts, including source code, models, and datasets, each governed by distinct and sometimes conflicting regimes.

Specifically, OSS licenses regulate how software can be legally reused, distributed, and modified, requiring downstream users to comply with license obligations to avoid legal risks like copyright infringement~\cite{11025699,li2025open}. Licensing challenges are already well-documented in traditional software; for example, Xu et al. report that 72.91\% of the 1,846 GitHub projects they studied exhibit license incompatibility~\cite{xu2023lidetector}. LLMware, as an emerging software paradigm with significantly more complex supply chains, is likely to amplify such challenges, yet its licensing risks remain underexplored. 
In particular, it remains unclear how traditional OSS licenses (e.g., \textsc{MIT} and \textsc{Apache-2.0}) interact and potentially conflict within LLMware supply chains, where OSS artifacts, LLMs, and datasets are tightly intertwined. Moreover, LLMware introduces licenses specific to LLMs and datasets (e.g., \textsc{OpenRAIL} and \textsc{LLaMA2}), giving rise to novel forms of incompatibility. While several prior studies have examined licensing issues either in OSS artifacts~\cite{mathur2012empirical,xu2023liresolver,xu2023lidetector,xu2025first,liu2026towards} or in LLM–dataset relationships~\cite{duan2024modelgo,yang2025ecosystem}, these works view LLM and source code components in isolation. The holistic licensing landscape of LLMware, shaped by heterogeneous dependencies, remains largely unexplored, motivating our study of license compatibility risks and legal compliance specific to LLMware.

In this paper, we first curate a large-scale dataset that captures the supply chain of LLMware by mining two major platforms: GitHub [\includegraphics[height=0.8em]{figs/github-mark.png}], the largest hub for OSS, and Hugging Face [\includegraphics[height=1em]{figs/hf-logo.png}], the most prominent platform for LLMs and datasets. Specifically, we identify GitHub repositories that depend on Hugging Face LLMs by matching 363 API signatures from 23 Python libraries maintained by Hugging Face (e.g., \textsc{transformers} and \textsc{diffusers}) via Sourcegraph~\cite{sourcegraph}, a large-scale code search engine. We then apply static code analysis to parse API invocations and extract LLM identifiers from function arguments, thus identifying the LLMs that each repository relies on. We iteratively trace the dependencies of these LLMs to their base models and training datasets using Hugging Face metadata, following a snowballing approach until no new dependencies are discovered. The resulting dataset comprises 12,180 GitHub repositories, 3,988 Hugging Face LLMs, and 708 datasets. For each component, we record its declared license if available or mark it as unlicensed otherwise. Based on this large dataset, we formulate four research questions (RQs) to systematically investigate licensing practices and challenges in the LLMware supply chain.

We first analyze the distribution of licenses in current LLMware supply chains (\textbf{RQ1}). Specifically, we quantify the license types declared across the three categories of artifacts in LLMware supply chains: OSS repositories, LLMs, and datasets, to characterize the overall licensing landscape. Our results reveal that \textsc{MIT} (4,475 occurrences, 26.5\%) and \textsc{Apache-2.0} (3,748 occurrences, 22.2\%) are the most prevalent licenses, while 5,983 artifacts (35.4\%) lack any license declaration. We further compare license distributions statistically and reveal three findings: (1) license practices differ significantly between GitHub and Hugging Face, with GitHub repositories predominantly adopting traditional OSS licenses while Hugging Face models adopt a more diverse set of licenses, including AI-specific ones (e.g., \textsc{OpenRAIL} and \textsc{LLaMA}); (2) company-released models more frequently adopt proprietary AI licenses, whereas individual developers have higher rates of missing licenses (339 vs. 902 no-license occurrences); and (3) license preferences vary across AI tasks, with LLMware specialized in Natural Language Processing (NLP) and Multimodal showing the widest variety.

We then investigate what aspects of licensing concern LLMware developers in practice (\textbf{RQ2}). We conduct a manual analysis of 337 license-related issues from GitHub repositories, 171 discussions from Hugging Face model repositories, and 84 discussions from Hugging Face dataset repositories within our collected LLMware supply chain, and perform open card sorting to build a taxonomy of developer concerns. Our analysis identifies seven categories, with license creation (54\%) and license update (30\%) being the most prevalent, together accounting for 84\% of discussions. This suggests that selecting and maintaining appropriate licenses remains the primary pain point for LLMware developers. Notably, license inquiries are substantially more common for LLMs (21.1\%) than for OSS (3.6\%) and datasets (9.6\%), indicating greater uncertainty about AI-specific license terms. We also observe notable differences in issue resolution: GitHub issues are resolved rapidly (61\% closed within one day), while Hugging Face discussions take significantly longer, with 50\% of LLM issues and 40\% of dataset issues remaining unresolved after two years. These findings highlight an urgent need for better license management guidance and tooling support in the LLMware ecosystem.


We further analyze license conflicts along LLMware supply chains (\textbf{RQ3}). We extract license terms and their corresponding attitudes (\textit{can}, \textit{cannot}, \textit{must}), model the supply chain as a directed graph, and detect conflicts between adjacent artifacts by comparing their license term attitudes, where a conflict is flagged when a downstream license is more permissive than its upstream counterpart (e.g., an upstream \textit{cannot} paired with a downstream \textit{can}). Our results reveal that license conflicts are pervasive, with 52\% of supply chains exhibiting at least one conflict across all three types of dependencies\footnote{$\mapsto$ denotes dependency; $\rightarrow$ denotes license conflict. Both arrows point from downstream to upstream.}: OSS $\mapsto$ LLM, LLM $\mapsto$ Dataset, and LLM $\mapsto$ Base Model. The most common conflict patterns involve: (1) missing licenses (e.g., No License$\rightarrow$\textsc{Apache-2.0}, 23.5\% in OSS$\rightarrow$LLM; \textsc{Apache-2.0}$\rightarrow$No License, 25.6\% in LLM$\rightarrow$Dataset); (2) incompatibilities between permissive OSS licenses due to differences in patent grants and attribution requirements, (e.g., \textsc{MIT}$\rightarrow$\textsc{Apache-2.0}, 15.1\%); and (3) mismatches between OSS licenses and content or AI-specific licenses that impose restrictions incompatible with permissive software reuse (e.g., \textsc{Apache-2.0} $\rightarrow$ \textsc{CC-BY-4.0}, 13.3\%; \textsc{Apache-2.0} $\rightarrow$ \textsc{LLaMA2}, 4.08\%). These conflicts highlight the complexity of managing licenses across heterogeneous artifacts in LLMware supply chains.

Finally, we evaluate existing license analysis approaches on detecting license conflicts in LLMware supply chains (\textbf{RQ4}). We construct a benchmark comprising 124 OSS licenses, 16 AI-specific licenses, and their mutated variants (620 OSS-Mut and 176 AI-Mut, generated by flipping one license term attitude at a time among \textit{can}, \textit{cannot}, and \textit{must}). Existing approaches show limited effectiveness: semantic similarity with sentiment analysis~\cite{karvelis2018topic, socher2013recursive} achieves 61\% F1 on OSS and 42\% on AI-specific licenses, while LiDetector~\cite{xu2023lidetector}, the state-of-the-art approach, achieves 76\% and 81\% respectively. We thus propose \ourtool, an LLM-based multi-agent framework for license conflict detection that employs an extraction agent to identify license terms and attitudes, followed by a repair agent that iteratively resolves inconsistencies. \ourtool achieves 88\% F1 on OSS and 89\% on AI-specific licenses, outperforming LiDetector by up to 14\%, and maintains strong performance on mutated licenses (86\% on OSS-Mut, 88\% on AI-Mut), demonstrating robustness to attitude perturbations.

To validate the practical impact of our findings, we report 60 license incompatibility issues detected by \ourtool to real-world LLMware projects. At the time of writing, 11 of 60 reported issues have been confirmed by developers, with 8 already fixed. Notably, 6 confirmed issues involve AI-specific licenses such as \textsc{LLaMA2} and \textsc{OpenRAIL++}.
Two LLMs with license conflicts are highly influential, with more than 107 million and 5 million downloads on Hugging Face, respectively.
These results suggest that such incompatibilities may impact a large number of downstream projects and users, and that \ourtool can effectively uncover practical and previously unnoticed license risks in the LLMware supply chain.


The main contributions of this paper include:
\begin{enumerate}[leftmargin=*]
    \item \textit{Dataset}: We curate a large-scale dataset capturing LLMware supply chains, including 12,180 GitHub repositories, 3,988 Hugging Face LLMs, and 708 datasets with their license and dependency information.
    \item \textit{Analysis}: We conduct a comprehensive empirical study on licensing practices in LLMware, revealing significant differences in license distributions and developer concerns across different supply chain artifacts (OSS, LLMs, and datasets), and license conflicts in 52\% of cases.
    \item \textit{Approach}: We propose \ourtool, an LLM-based multi-agent framework for license compatibility analysis, achieving up to 87\% F1 in detecting license conflicts and outperforming prior state-of-the-art by up to 14\%.
    \item \textit{Practices}: We report 60 license issues to real-world projects, with 11 confirmed (including two involving LLMs with over 500 million downloads), and derive actionable lessons for developers.
\end{enumerate}

\section{Background and Related Work}

\begin{figure}[tp!]
    \centering
    \includegraphics[width=\linewidth]{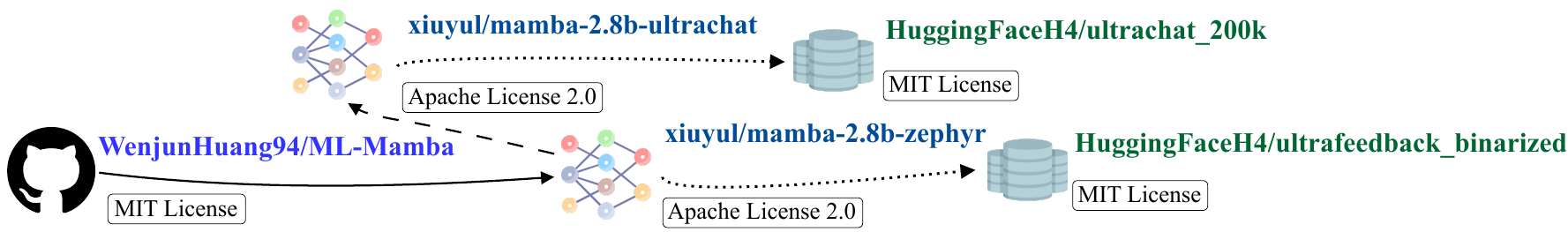}
    \caption{The example supply chain of LLMware}
    \figlabel{supply-chain}
\end{figure}

\subsection{LLMware and Its Supply Chains}

Large Language Models (LLMs) have achieved remarkable performance across diverse tasks such as natural language dialogue~\cite{10.1145/3771090,10.1145/3641289}, program synthesis~\cite{10.1145/3695988,10.1145/3708525}, and even autonomous driving~\cite{10495699,10629039}. Motivated by these advances, technology companies and individual developers have increasingly built LLM-powered applications~\cite{10.1145/3708531}, leading to a new class of intelligent software that embeds or invokes LLMs as core components. We refer to this class as \textit{LLMware}, with prominent examples including ChatGPT~\cite{chatgpt} and GitHub Copilot~\cite{copilot}.

Similar to traditional software, LLMware development involves multiple stakeholders and artifacts, such as model developers, code contributors, and data providers~\cite{10.1145/3708531,10.1145/3713081.3731747}. The integration of these components forms a software supply chain encompassing all upstream artifacts and dependencies that a software system relies on. However, LLMware supply chains differ fundamentally from traditional ones: while conventional supply chains primarily consist of OSS libraries and packages, LLMware introduces additional dependencies on LLMs, which may further rely on base models, for example through fine-tuning~\cite{yang2025ecosystem} or distillation~\cite{10.1145/3708525}, as well as training datasets. Consequently, LLMware supply chains span three types of heterogeneous artifacts: OSS repositories, LLMs, and datasets. \figref{supply-chain} illustrates an example of such a supply chain, where the OSS project `WenjunHuang94/ML-Mamba' on GitHub depends on two Hugging Face models, `xiuyul/mamba-2.8b-ultrachat' and `xiuyul/mamba-2.8b-zephyr', which are further trained on Hugging Face datasets `ultrachat\_200k' and `ultrafeedback\_binarized'. Although recent studies have examined various aspects of the LLM ecosystem~\cite{10.1145/3708531,10.1145/3713081.3731747,yang2025ecosystem,duan2024modelgo}, they largely focus on individual components in isolation. In contrast, our work adopts a holistic view to investigate licensing practices across all three artifact types, which we believe is novel and offers a unique perspective in this field.

\subsection{License Analysis}

As introduced in Section~\ref{sec:intro}, software licenses regulate how software can be legally reused, distributed, and modified, and they require users to comply with license obligations to avoid copyright risks~\cite{11025699,li2025open}. However, ensuring license compliance across software supply chains is challenging, and license conflicts can easily arise. In LLMware, where multiple artifact types with distinct licenses are tightly coupled, \figref{supply-chain} illustrates a concrete example of a license conflict. The GitHub project `WenjunHuang94/ML-Mamba' is released under the \textsc{MIT} license and depends on two Hugging Face LLMs licensed under \textsc{Apache-2.0}, which are trained on datasets distributed under the \textsc{MIT} license. Because \textsc{MIT} and \textsc{Apache-2.0} contain incompatible terms related to patent grants and attribution requirements~\cite{xu2023lidetector}, putting them within a single LLMware can introduce potential legal risks.

Extensive studies have examined license detection and compliance in traditional OSS. The Software Package Data Exchange (SPDX)~\cite{stewart2010software,mancinelli2006managing} provides a standardized format for license information and is recognized as an international open standard (ISO/IEC 5962:2021)~\cite{spdx}. Building on this, researchers have proposed automated approaches for license compatibility analysis~\cite{kapitsaki2017automating,liu2024catch,xu2023lidetector,xu2023liresolver} and developed tools for license identification~\cite{german2010sentence,gobeille2008fossology,kapitsaki2017identifying}. There are also empirical studies have characterized licensing practices across OSS ecosystems. Mathur et al.~\cite{mathur2012empirical} identified license violations in Google Code projects, while Wu et al.~\cite{wu2024large} revealed license irregularities across package managers. Other studies examined developer confusion about licenses~\cite{papoutsoglou2022analysis}, the impact of license changes~\cite{han2026cloud}, and license compliance in LLM-generated code~\cite{xu2024licoeval}. Li et al.~\cite{li2025open} provided a comprehensive survey classifying research into license identification, risk assessment, and mitigation.
Recent work has begun addressing licensing in AI ecosystems. Yang et al.\cite{yang2025ecosystem} analyzed licensing practices among models and datasets on Hugging Face, while Duan et al.\cite{duan2024modelgo} proposed ModelGo for auditing legal risks in ML projects. Other efforts include detecting license violations during model forking~\cite{huang2024your}, exploring LLMs for license conflict analysis~\cite{cui2023empirical,cui2025exploring,ke2025clausebench}, and studying model reuse practices~\cite{jiang2023empirical}.
However, existing work largely treats software, LLMs, and datasets in isolation. Our work bridges this gap by examining the holistic licensing landscape of LLMware, where heterogeneous dependencies across all three artifact types create novel compatibility challenges.
\section{LLMware Supply Chain Construction}
Here we describe how we construct the LLMware supply chain dataset for our empirical analysis. Our goal is to establish dependency mappings across three artifact types: OSS repositories, LLMs, and datasets. The construction involves multiple steps: searching GitHub repositories that leverage Hugging Face's officially maintained libraries, matching API signatures to identify the LLMs each repository invokes, and tracing to base models and training datasets via metadata analysis. The following subsections detail these steps.

\subsection{Bridging GitHub Repositories to Hugging Face LLMs}
\label{sec:mapping}

Software is primarily composed of code, and so is LLMware. On GitHub, numerous open-source projects leverage LLMs as core components, either by directly embedding models or invoking them via remote APIs. These projects represent the most downstream artifacts in our defined LLMware supply chain, serving as end-user applications that depend on upstream LLMs~\cite{10.1145/3708531,10.1145/3713081.3731747}. However, GitHub is not the primary platform for hosting and sharing LLMs; Hugging Face serves that role~\cite{jiang2024peatmoss,10173952,yang2025ecosystem,duan2024modelgo}. Therefore, to analyze license relationships across software, models, and datasets, we first establish mappings between the GitHub and Hugging Face communities.
Specifically, we mine repositories from GitHub (the largest hub for OSS) and Hugging Face (the most prominent hub for LLMs and datasets). We adopt the data collection procedure of PeaTMOSS~\cite{jiang2024peatmoss} to map OSS repositories to their dependent models, omitting PyTorch Hub due to its limited scale. We then trace each model to its base models and training datasets. The detailed steps are as follows.

\smalltitle{Mining OSS repositories via API Signatures}
The first challenge lies in identifying LLMware from GitHub, i.e., OSS projects that depend on LLMs hosted on Hugging Face. LLMs on Hugging Face predominantly follow the Transformer architecture~\cite{vaswani2017attention} and are accessed via libraries officially maintained by Hugging Face~\cite{jiang2024peatmoss,10173952,yang2025ecosystem}, such as \textsc{transformers}, \textsc{diffusers}, and \textsc{peft}. OSS projects typically load these models by calling library functions with model identifiers as arguments (e.g., `AutoModel.from\_pretrained("bert-base-uncased")'), forming standardized API usage patterns. PeaTMOSS~\cite{jiang2024peatmoss}, an early work analyzing Hugging Face models, identifies these patterns as \textit{signatures} of LLM usage, which consist of import statements (e.g., `import transformers' or `from diffusers import StableDiffusionPipeline') that suggests dependencies on Hugging Face libraries. We reuse the manually curated signatures from PeaTMOSS and similarly focus exclusively on Python, as it is the dominant language for interacting with LLMs~\cite{jiang2024peatmoss,yang2025ecosystem}. In total, we use 363 signatures from 23 Python libraries that access Hugging Face.

For each API signature, we search using Sourcegraph~\cite{sourcegraph}, a large-scale code search engine that indexes public GitHub repositories with $\geq$5 stars. Sourcegraph enables precise and scalable search over vast amounts of open-source code and prioritizes repositories with higher popularity through star-based filtering, returning representative and up-to-date results. Each search result includes the repository name, source file path, and matched code snippet. We conducted our search in February 2025, with no earlier cutoff date, capturing all historical repositories matching our criteria. The identified GitHub repositories as well as their declared license are recorded for subsequent analysis.

\vspace{0.1cm}
\noindent \textbf{Static Code Analysis}
We develop customized static analysis using the Scalpel~\cite{li2022scalpel} framework to filter out false positives and identify actual API invocations. Specifically, we parse each matched Python file into an Abstract Syntax Tree (AST) and traverse it to locate function calls matching Hugging Face API patterns (e.g., `from\_pretrained()', `pipeline()'). For each identified call, we extract model identifiers from function arguments through constant propagation and string analysis, discarding repositories that contain only import statements without actual invocations or where identifiers cannot be resolved statically.

\begin{wrapfigure}{r}{0.47\linewidth}
    \centering
    \includegraphics[width=\linewidth]{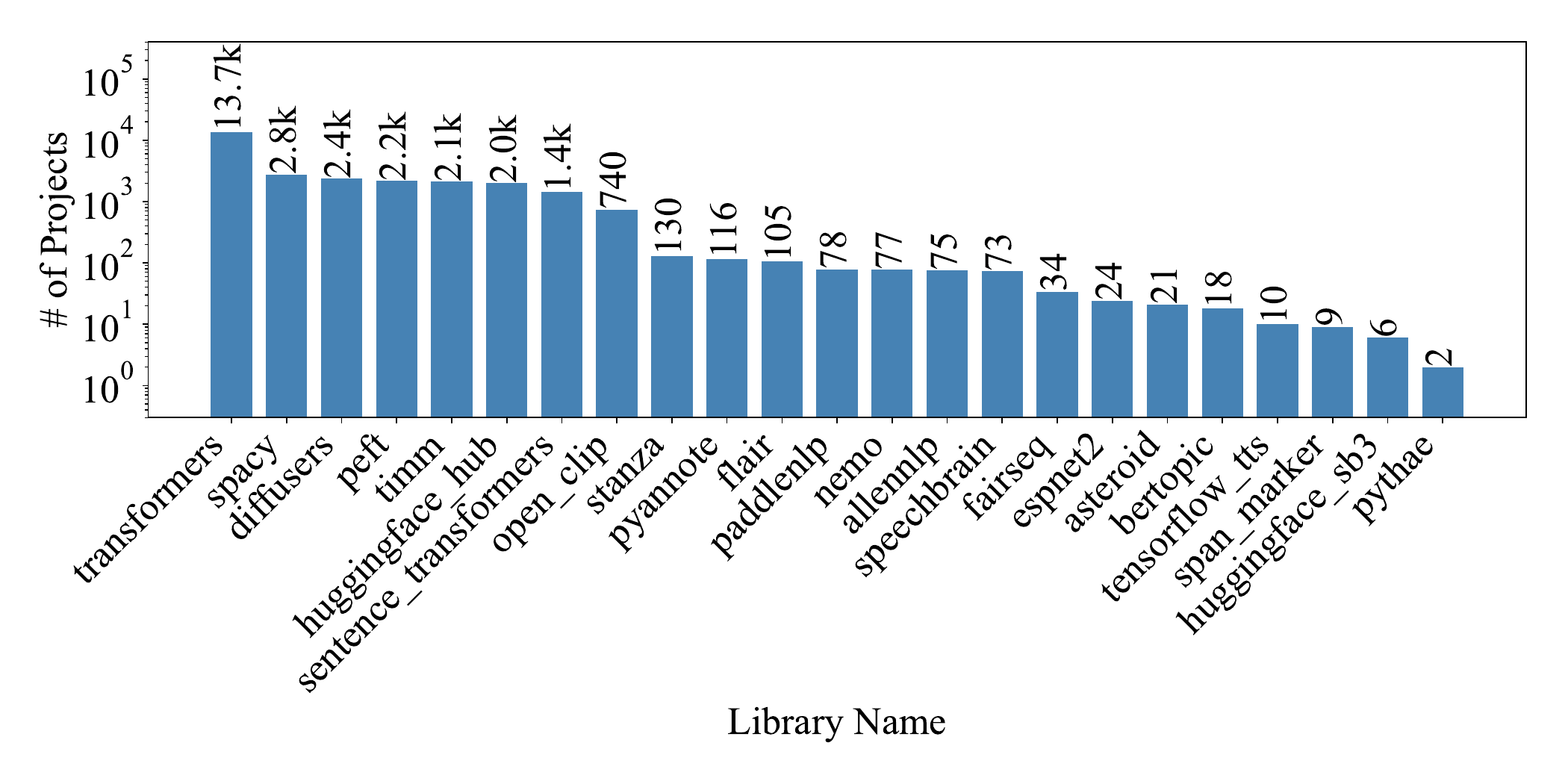}
    \caption{Number of projects accessing Hugging Face LLMs searched via Sourcegraph}
    \figlabel{proj-numbers}
\end{wrapfigure}

After removing forked repositories, we identified 28,065 GitHub repositories that correctly use these libraries. \figref{proj-numbers} shows the distribution of API usage across all 23 Hugging Face libraries. The five most frequently used are \textsc{transformers}, \textsc{spacy}, \textsc{diffusers}, \textsc{peft}, and \textsc{timm}, with \textsc{transformers} being significantly more prevalent. This distribution differs from PeaTMOSS~\cite{jiang2024peatmoss}, suggesting that library popularity evolves over time. For example, \textsc{diffusers} ranked fifth in PeaTMOSS but appears third in our results, indicating growing interest in image generation models.

\vspace{0.1cm}
\noindent \textbf{Mapping Hugging Face LLMs}
Using the filtered GitHub repositories and the program information extracted via the static analysis above, we map each repository to the Hugging Face LLMs it depends on. As described earlier, we extract model identifiers from function arguments (e.g., the string `bert-base-uncased' in `from\_pretrained("bert-base-uncased")') for each API invocation identified by our static analysis. Unlike generic code search engines that may match comments or dead code, Scalpel retains only actual code-level invocations~\cite{li2022scalpel}, enabling reliable model mapping. We then query the Hugging Face Hub API to validate that each extracted identifier corresponds to an existing LLM, filtering out invalid or deprecated references.
The resulting mapping is many-to-many: a single repository may depend on multiple LLMs, and a single LLM may be used by multiple repositories. As a result, we identify 12,180 GitHub repositories that collectively depend on 3,988 distinct Hugging Face LLMs. To verify accuracy, we randomly sampled and manually inspected 50 mappings, all of which were correct.


\subsection{Bridging LLMs and Datasets}
With the GitHub repositories and Hugging Face models collected above, the next step is to connect LLMs with their training datasets. As core components of LLMware, LLMs sit in the middle of the supply chain: they rely on upstream datasets for training knowledge while providing intelligence capabilities to downstream software applications. Hugging Face serves as the largest hub for hosting both models and datasets, and provides comprehensive dependency information through its metadata system.
Specifically, Hugging Face provides \textit{metadata} for each model and dataset~\cite{huggingfaceModelCards}, which is used to render its corresponding webpage. This metadata includes key attributes such as the license type, associated training datasets if any, and the base model if any from which a given model was derived. 
\begin{wrapfigure}{r}{0.45\linewidth}
    \centering
    \includegraphics[width=\linewidth]{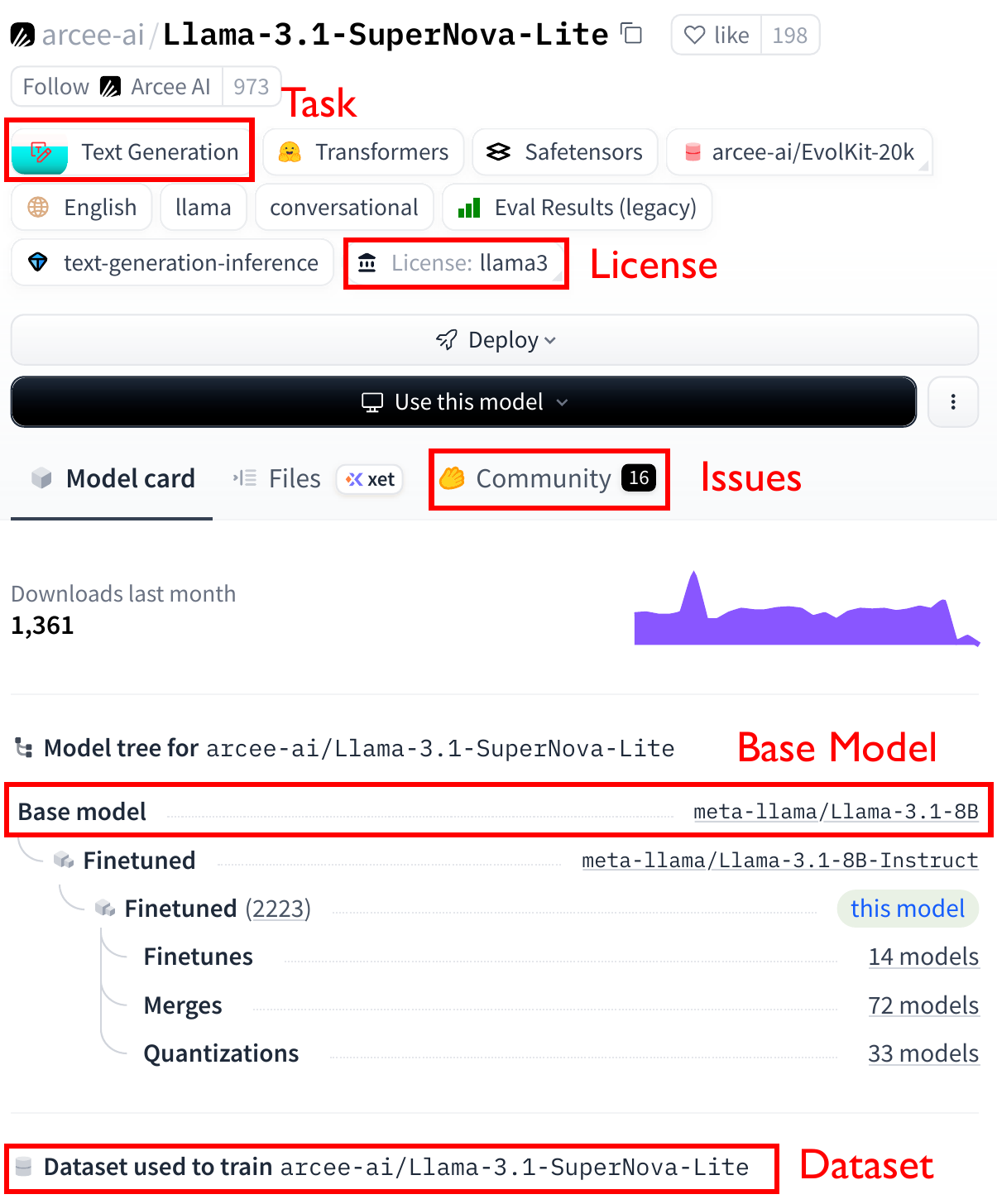}
    \caption{Example: Hugging Face model metadata}
    \figlabel{model-card}
\end{wrapfigure}
\figref{model-card} illustrates an example of a Hugging Face model card for the model `Llama-3.1-SuperNova-Lite'~\cite{ArceeaiLlama31}. Its metadata shows that the model is licensed under the AI-specific license \textsc{LLaMA3}, fine-tuned on the base LLM `Llama-3.1-8B-Instruct', and trained on the dataset `Llama-3.1-SuperNova-Lite'.

For each model retrieved in the previous step, we parse its metadata to extract key attributes, including its license, base model, and associated training datasets. Following a snowballing approach~\cite{wohlin2014guidelines,jalali2012systematic}, we start from the 3,988 LLMs identified above and query the Hugging Face API to retrieve each model's metadata. When a model references a base model or training dataset, we recursively fetch the metadata of these newly discovered artifacts until no new dependencies are found, collecting 708 datasets in total via this process. By combining all mappings, we construct a large-scale dataset of the LLMware supply chain capturing dependencies across three artifact types: 12,180 GitHub repositories, 3,988 LLMs, and 708 datasets, with each artifact's declared license recorded for subsequent analysis.





\section{RQs and Empirical Results}

With the constructed LLMware supply chain dataset, we formulate and investigate four research questions (RQs) to systematically analyze licensing practices and challenges:

\begin{itemize}[leftmargin=*]
    \item \textbf{RQ1:} What are the license distributions in current LLMware systems?
    \item \textbf{RQ2:} What aspects of licensing are LLMware developers concerned with in practice?
    \item \textbf{RQ3:} What kinds of license conflicts exist in LLMware supply chains?
    \item \textbf{RQ4:} To what extent can existing approaches detect license conflicts in LLMware?
\end{itemize}

\subsection{RQ1: License Distributions}

\subsubsection{Motivation}
Compared to traditional OSS, LLMware incorporates not only OSS components but also LLMs and their training datasets, resulting in a more complex and heterogeneous supply chain. Unlike traditional OSS ecosystems, where licensing practices are well-studied~\cite{xu2023lidetector,xu2023liresolver,li2025open}, the licensing landscape of LLMware remains largely unexplored. Key questions include: How do license distributions differ between GitHub OSS projects and Hugging Face models/datasets? Do company-released models exhibit different licensing patterns compared to individual developers? Are there domain-specific licensing preferences across AI task categories? Answering these questions is essential for understanding compliance risks in LLMware and providing empirical evidence to guide license selection practices.

\subsubsection{Methodology}
Based on the constructed supply chain, we perform a quantitative analysis of license distributions across different artifact types. We extract license information for each artifact and calculate the frequency and proportion of each license type in GitHub repositories, Hugging Face LLMs, and datasets. We further group LLMs by task categories (e.g., NLP, computer vision) and owner types (company vs. individual) for comparative analysis. To test whether distributions differ significantly, we conduct Chi-square tests~\cite{doi:10.1177/1098214011426594} and measure effect sizes using Cramér's V~\cite{cramér1946mathematical}, where V $<$ 0.10 indicates negligible, 0.10--0.20 weak, 0.20--0.40 moderate, and $>$ 0.40 strong association.

\begin{wraptable}{r}{0.42\linewidth}
  \centering
  \scriptsize
  \caption{The overall license frequency}
    \begin{tabular}{  l | l | l }
    \hline 
    \bf Rank & \bf License & \bf Count \\
    \hline \hline
    \bf 1st & \code{MIT} & 4475 \\ \hline
    \bf 2nd & \code{Apache-2.0} & 3748 \\ \hline
    \bf 3rd & \code{GPL-3.0} & 382 \\ \hline
    \bf 4th & \code{BSD-3-Clause} & 224 \\ \hline
    \bf 5th & \code{CC-BY-4.0} & 151 \\ \hline
    \hline 
    - & \code{No License} & 5983 \\
    \hline
    - & \code{Others} & 945 \\
    \hline
    \end{tabular}
    \tablabel{overall-frequency}
\end{wraptable}

\subsubsection{Results}

We extract license information for each artifact collected from GitHub and Hugging Face for this analysis. For rare licenses, specifically those not included in GitHub’s recognized license list, both platforms label them as \code{Others}, while artifacts without a declared license are categorized as \code{No License}. \tabref{overall-frequency} presents the top five most frequently used licenses, together with counts for \code{Others} and \code{No License}. The most common licenses are \code{MIT} (4,475, 26.5\%) and \code{Apache-2.0} (3,748, 22.2\%), followed by \code{GPL-3.0}, \code{BSD-3-Clause}, and \code{CC-BY-4.0}. Notably, 5,983 artifacts (35.4\%) have no license declaration, while 945 artifacts (5.6\%) use less common licenses. These results indicate a strong concentration in a few permissive licenses, alongside a substantial portion of artifacts with unclear licensing terms.

To further analyze license distributions in LLMware, we formulate three hypotheses:
\begin{itemize}[leftmargin=*]
    \item \textbf{Hypothesis 1:} \textit{The distribution of licenses on GitHub is consistent with that on Hugging Face.}
    \item \textbf{Hypothesis 2:} \textit{The license distribution of company-released models does not significantly differ from that of individual-released models.}
    \item \textbf{Hypothesis 3:} \textit{The license distributions across different AI task categories are statistically similar.}
\end{itemize}

\smalltitle{License distributions across different artifact types}
LLMware comprises OSS projects from GitHub and models/datasets from Hugging Face, developed by diverse contributors under different licensing practices. Analyzing license distributions across these artifact types helps clarify potential compliance risks and highlight differences in licensing norms. \figref{overall-distribution} shows the license distributions for GitHub OSS projects and Hugging Face models/datasets. For OSS projects (\figref{oss-repos}), traditional licenses such as \code{MIT} and \code{Apache-2.0} dominate. In contrast, Hugging Face models and datasets (\figref{hf-llm-dataset}) display a more diverse landscape, with AI-specific licenses such as \code{BigScience RAIL} and \code{OpenRAIL-M} accounting for a substantial share. Notably, \code{No License} is the most common category on both platforms, raising concerns about reuse ambiguity and compliance risks.

\begin{figure}[t]
  \centering
  \begin{subfigure}[t]{0.49\linewidth}
    \centering
    \includegraphics[width=\linewidth]{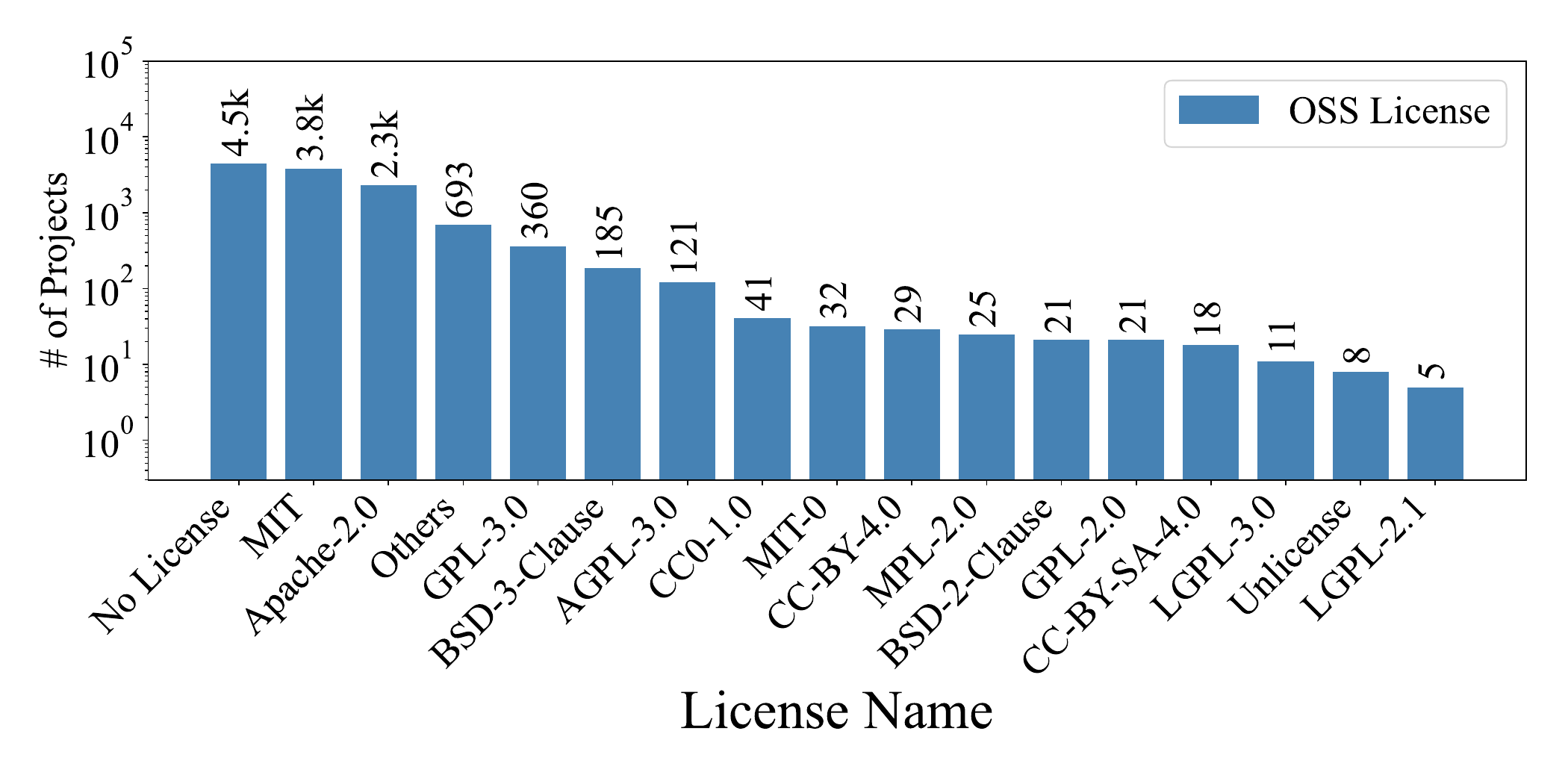}
    \caption{GitHub repositories}
    \figlabel{oss-repos}
  \end{subfigure}
  \hfill
  \begin{subfigure}[t]{0.49\linewidth}
    \centering
    \includegraphics[width=\linewidth]{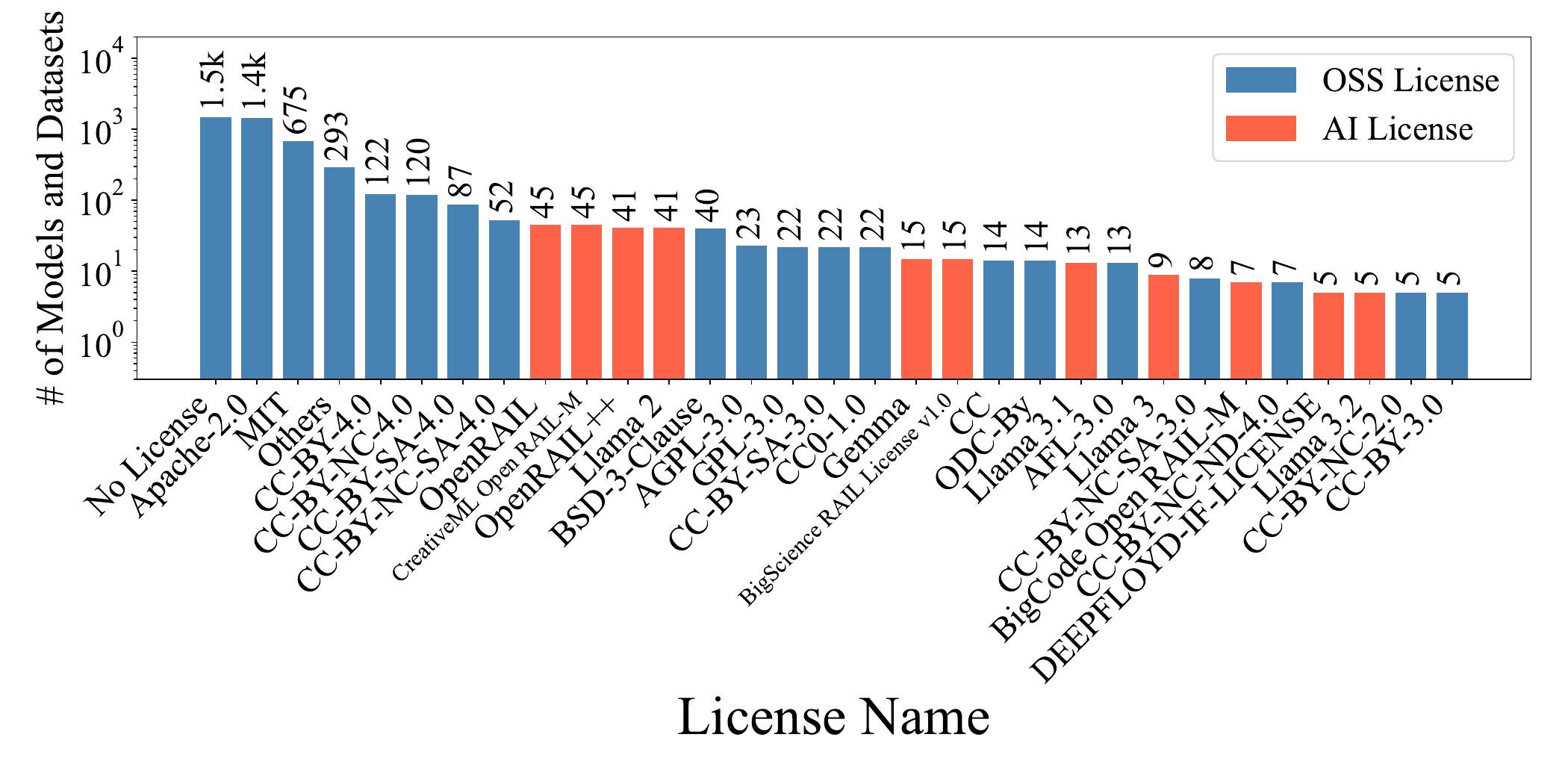}
    \caption{Hugging Face LLMs and datasets}
    \figlabel{hf-llm-dataset}
  \end{subfigure}
  \caption{The overall license distribution of LLMware components}
  \figlabel{overall-distribution}
\end{figure}

To test \textbf{Hypothesis 1}, we conduct a Chi-square test to compare license distributions between the two platforms. The results indicate a statistically significant difference ($p < 0.001$) with Cramér's V of 0.37 (medium-to-large effect), leading us to reject the null hypothesis. This confirms that license practices differ significantly between GitHub and Hugging Face, challenging traditional compatibility assumptions and highlighting the need to re-evaluate license analysis approaches designed for traditional OSS.


\smalltitle{License distributions across different owner types}
We classify LLMs into \textit{company models} and \textit{individual models} based on Hugging Face metadata tags (e.g., \textit{company}, \textit{enterprise}); models without such tags are considered individual models. \figref{all-owners} compares the license distributions between company models (\figref{company-models}) and individual models (\figref{individual-models}). Both groups predominantly adopt OSS licenses such as \code{Apache-2.0} and \code{MIT}, with \code{Apache-2.0} being the most common (509 company vs. 816 individual models). However, company owners tend to declare licenses more frequently, while individual owners have a higher proportion of \code{No License} (902). AI-specific licenses are also more common among company models: proprietary licenses such as \code{LLaMA2}, \code{LLaMA3}, and \code{OpenRAIL++} appear more frequently, while individual developers favor community-oriented licenses like \code{CreativeML Open RAIL-M} and \code{BigScience RAIL}. Testing \textbf{Hypothesis 2} with a Chi-square test confirms a statistically significant difference ($p < 0.001$, Cramér's V = 0.26, moderate effect). These findings suggest that companies adopt more formal licensing strategies tied to commercialization, while individuals prioritize flexibility or simply omit declarations.

\begin{figure}[t]
  \centering
  \begin{subfigure}[t]{0.49\linewidth}
    \centering
    \includegraphics[width=\linewidth]{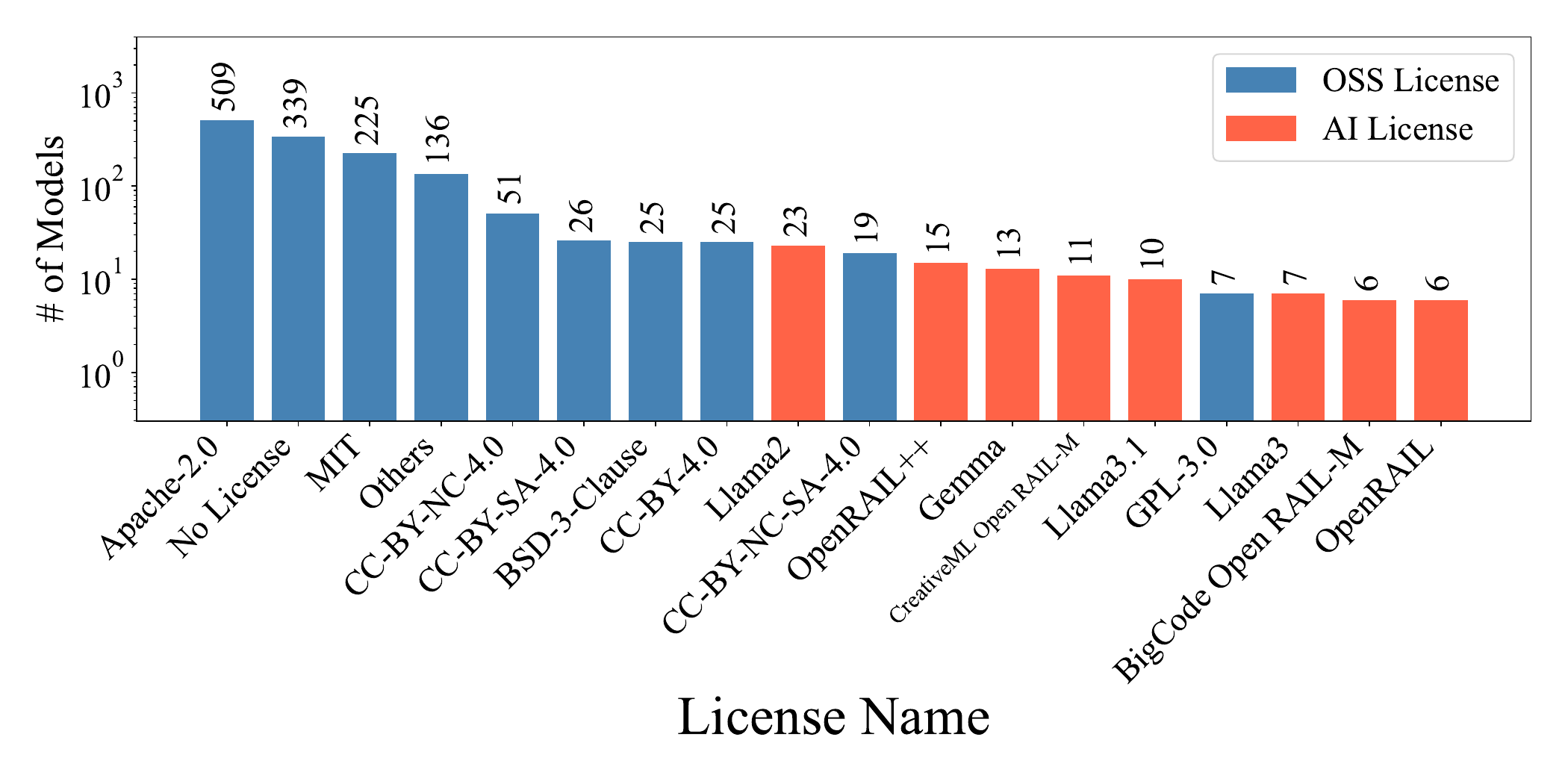}
    \caption{Company LLMs}
    \figlabel{company-models}
  \end{subfigure}
  \begin{subfigure}[t]{0.49\linewidth}
    \centering
    \includegraphics[width=\linewidth]{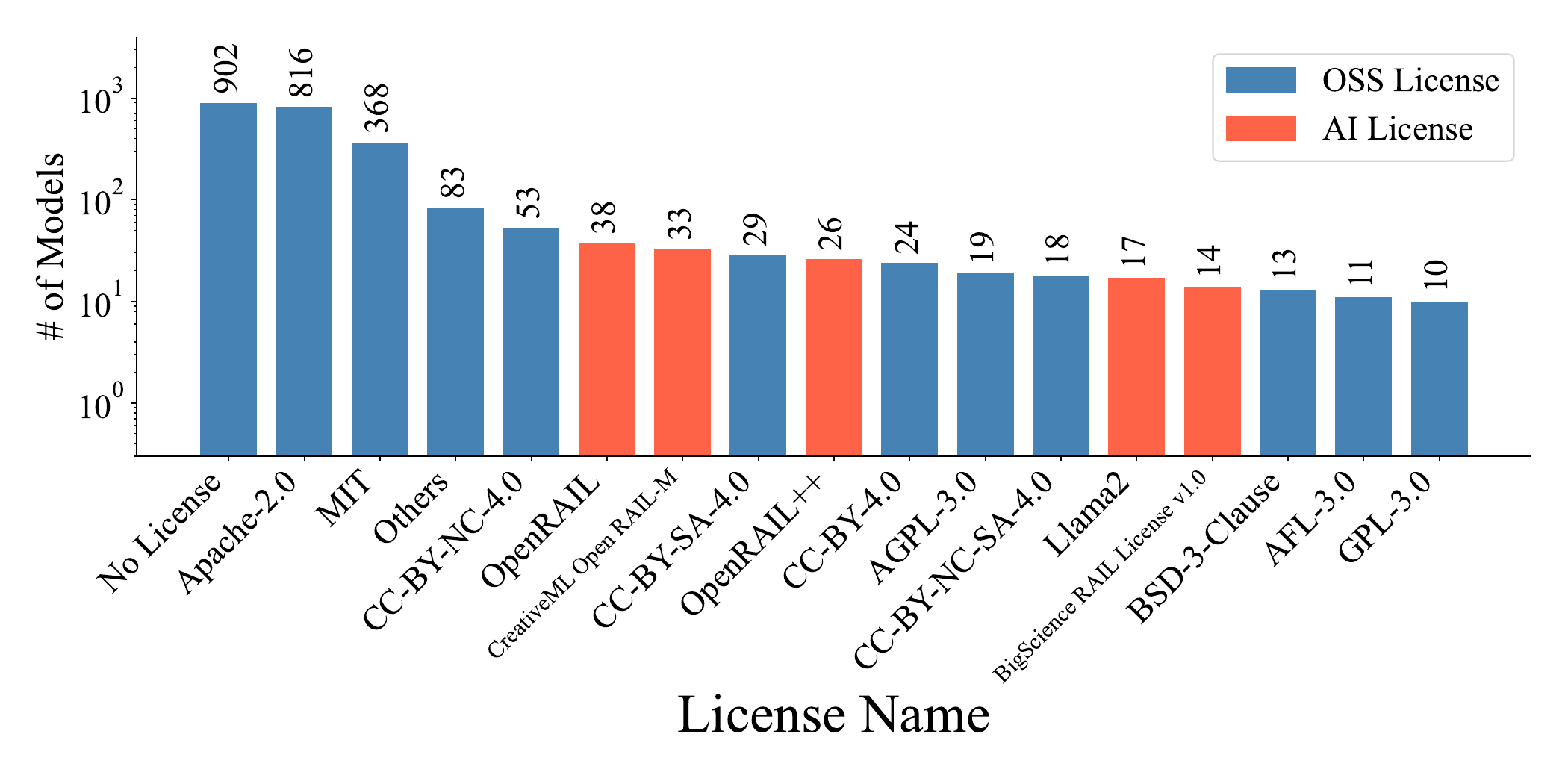}
    \caption{Individual LLMs}
    \figlabel{individual-models}
  \end{subfigure}
  \caption{The license distributions of different owners}
  \figlabel{all-owners}
\end{figure}

\begin{figure}[t!]
  \centering
  \begin{minipage}{0.42\linewidth}
    \centering
    \includegraphics[width=\linewidth]{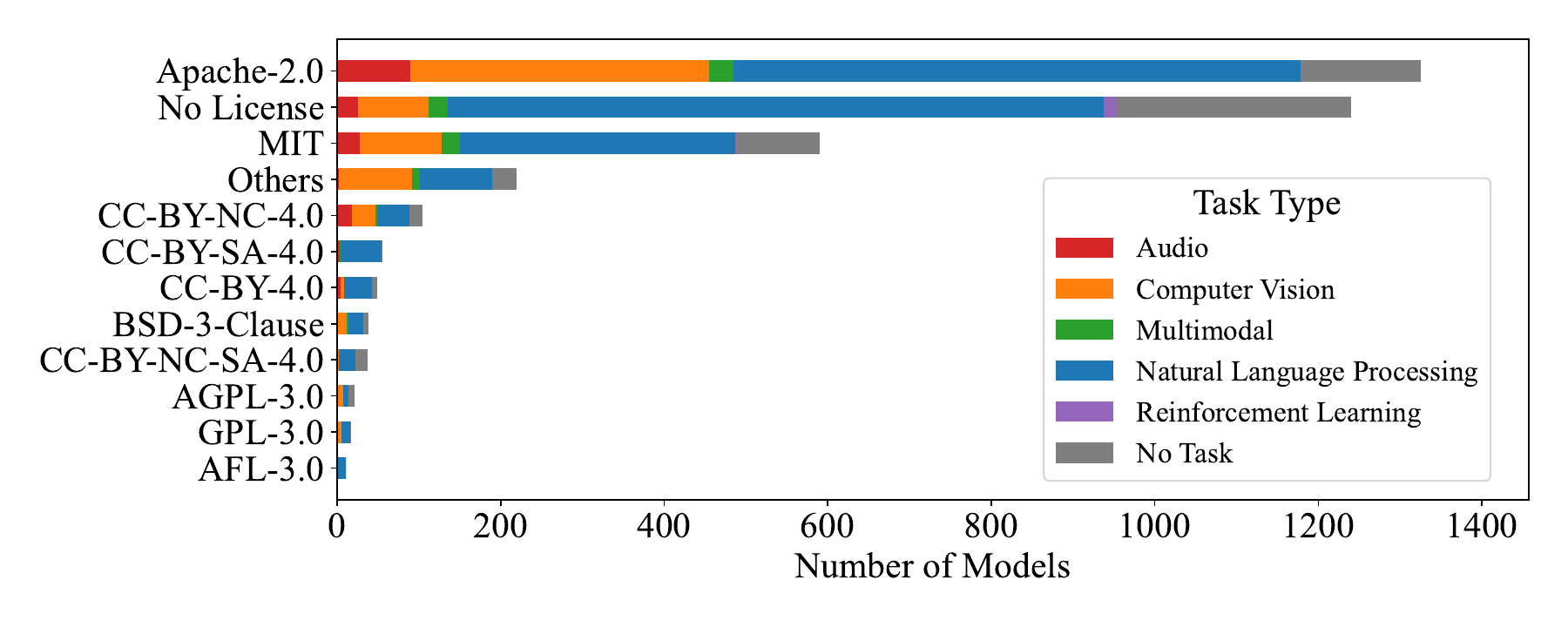}
    \subcaption{OSS licenses across tasks.}
    \label{fig:oss-tasks}
  \end{minipage}
  \hfill
  \begin{minipage}{0.57\linewidth}
    \centering
    \includegraphics[width=\linewidth]{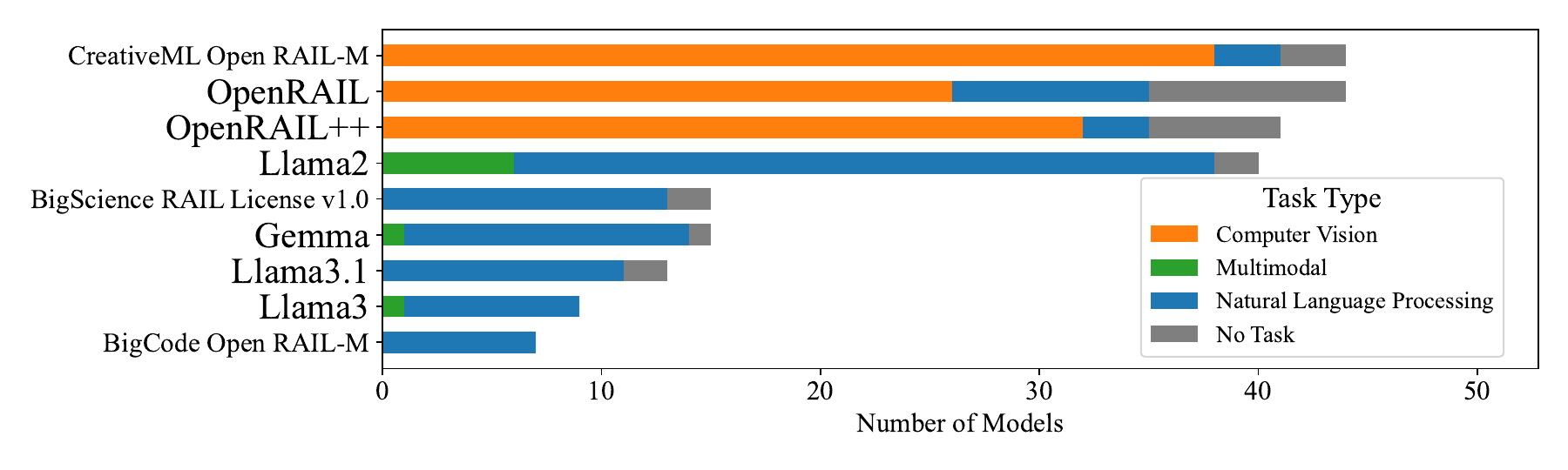}
    \subcaption{AI-specific Licenses across tasks.}
    \label{fig:ai-tasks}
  \end{minipage}
  \caption{The license distributions of Hugging Face LLMs across tasks.}
  \label{fig:all-tasks}
\end{figure}


\smalltitle{License distributions across different task categories}
We classify LLMs by task categories defined by Hugging Face: \textit{Audio}, \textit{Computer Vision (CV)}, \textit{Multimodal}, \textit{Natural Language Processing (NLP)}, \textit{Reinforcement Learning}, \textit{Tabular}, and \textit{Other}. LLMs without specified tasks are labeled as \textit{No Task}. \figref{all-tasks} presents license distributions across task categories, separated by OSS licenses (\figref{oss-tasks}) and AI-specific licenses (\figref{ai-tasks}). For OSS licenses, \textit{NLP} tasks dominate the usage of popular licenses like \code{Apache-2.0} and \code{MIT}, while \textit{Audio}, \textit{CV}, and \textit{Multimodal} tasks use a narrower range. AI-specific licenses show even greater task-specific concentration: \code{OpenRAIL}, \code{OpenRAIL++}, and \code{CreativeML Open RAIL-M} are most common in \textit{Multimodal} and \textit{NLP} models, while proprietary licenses like \code{LLaMA2}, \code{LLaMA3}, and \code{Gemma} are almost exclusively tied to \textit{NLP} models.

To test \textbf{Hypothesis 3}, we conduct pairwise Chi-square tests between task categories. Most pairs show statistically significant differences ($p \leq 0.003$, Cramér's V $\geq$ 0.21, moderate effect), except for \textit{NLP} vs. \textit{Multimodal} ($p = 0.261$, V = 0.13) and \textit{NLP} vs. \textit{Reinforcement Learning} ($p = 0.424$, V = 0.10), which do not differ significantly. These findings suggest that license distributions vary across most task categories, while \textit{NLP}, \textit{Multimodal}, and \textit{Reinforcement Learning} share similar licensing practices, likely due to their common reliance on large-scale pretrained models and shared licensing norms in the generative AI community.


\finding{License distributions vary significantly across artifact types, owner types, and task categories in LLMware. Over one-third of artifacts lack declarations, and AI-specific licenses are prevalent on Hugging Face, challenging traditional OSS compatibility assumptions.}

\subsection{RQ2: Concerns of Developers}

\subsubsection{Motivation}
LLMware is built on complex dependencies involving OSS, LLMs, and datasets, each potentially governed by distinct licenses. In practice, managing these licenses is not only a technical necessity but also a community-driven concern. Developers often discuss licensing issues through GitHub's issue trackers and Hugging Face's community forums, as well as the pull request features available on both platforms, reflecting real-world challenges in ensuring legal compliance. Understanding what aspects of licensing developers are most concerned about helps identify pain points in license management, such as selecting compatible licenses, resolving conflicts, or understanding AI-specific license implications.

\subsubsection{Methodology}
To understand developers' practical concerns regarding licenses in LLMware, we analyze real-world discussions from GitHub and Hugging Face. We implement a crawler to collect license-related issues from GitHub repositories and discussions from Hugging Face model and dataset pages, filtering by license-related keywords (e.g., ``license'', ``MIT'', ``GPL'', ``Apache-2.0''). In total, we collect 2,613 GitHub issues from 1,301 repositories (11\% of projects in our dataset), 307 discussions from 224 LLMs (6\%), and 106 discussions from 83 datasets (12\%). For in-depth analysis, we randomly sample 384 GitHub issues, 171 LLM discussions, and 84 dataset discussions using a 95\% confidence level with a 5\% margin of error. After manual validation to eliminate irrelevant entries, 337 GitHub issues (88\%) are confirmed as license-related, while all 255 Hugging Face discussions (171 + 84) are retained.

We adopt an open card-sorting approach~\cite{yang2025ecosystem} to develop a taxonomy of developers' licensing concerns. Two annotators (graduate students with at least 1.5 years of SE research experience) and one senior researcher (15+ years of experience) first jointly reviewed 11\% of the samples (approximately 60 discussions) to derive initial categories without predefined themes, allowing patterns to emerge naturally from the data. Based on these discussions, we finalized a codebook with category definitions, inclusion/exclusion criteria, and representative examples (available in our replication package). The two annotators then independently labeled the remaining samples using the fixed codebook. No new categories emerged during this phase, suggesting that our taxonomy reached saturation. Disagreements were resolved through discussion, with the senior researcher adjudicating when consensus could not be reached.

Following prior empirical studies~\cite{macklon2023taxonomy,liu2023towards,wang2025comprehensive,humbatova2020taxonomy,tan2014bug}, we assess inter-rater reliability on the independently labeled portion using Cohen's Kappa ($\kappa$)~\cite{viera2005understanding}. We obtain $\kappa=0.83$ for GitHub issues, $\kappa=0.82$ for LLM discussions, and $\kappa=0.86$ for dataset discussions (overall $\kappa=0.83$), indicating substantial agreement between annotators.

\subsubsection{Results}

This labeling results in a taxonomy comprising \textit{seven} categories that represent the major aspects of license concerns in LLMware development, shown as follows:
\begin{enumerate}[leftmargin=*]
    \item \textbf{License Creation}: Adding a license to an unlicensed project, LLM, or dataset.
    \item \textbf{License Update}: Modifying an existing license due to inappropriateness or incompatibility.
    \item \textbf{License Inquiry}: Requesting clarification about license terms, e.g., whether commercial use or model modification is permitted.
    \item \textbf{License Selection Justification}: Discussing the rationale for choosing a specific license over alternatives.
    \item \textbf{License Conflict}: Raising concerns about conflicts between the current license and other licenses in the software system.
    \item \textbf{LLM License Compatibility}: Discussing compatibility between the project license and dependent LLMs.
    \item \textbf{Dataset License Compatibility}: Discussing compatibility between the project license and training datasets.
\end{enumerate}

\begin{table}[t!]
  \centering
  \footnotesize
  \caption{Distribution of license discussion}
    \begin{tabular}{l|l|l|l}
    \hline
    \textbf{Category} & \textbf{OSS} & \textbf{LLM} & \textbf{Dataset} \\
    \hline \hline
    \textbf{License Creation} & 195 (57.9\%) & 89 (52.0\%) & 37 (44.6\%)\\ \hline
    \textbf{License Update} & 101 (30.0\%) & 41 (24.0\%) & 36 (43.4\%)\\ \hline
    \textbf{License Inquiry} & 12 (3.6\%) & 36 (21.1\%) & 8 (9.6\%)\\ \hline
    \textbf{License Selection Justification} & 5 (1.5\%) & 3 (1.8\%) & 2 (2.4\%)\\ \hline
    \textbf{License Conflict} & 4 (1.2\%) & 1 (0.6\%) & 1 (1.2\%)\\ \hline
    \textbf{LLM License Compatibility} & 6 (1.8\%) & 0 (0\%) & - \\ \hline
    \textbf{Dataset License Compatibility} & 16 (4.2\%) & 1 (0.6\%) & 0 (0\%)\\
    \hline \hline
    \textbf{All} & 337 & 171 & 84 \\
    \hline
    \end{tabular}
  \tablabel{discussion}
\end{table}

\smalltitle{License-related Discussion Distribution}
As shown in \tabref{discussion}, we observe clear differences in the distribution of license-related discussions across OSS repositories, LLMs, and datasets. Most discussions in all three domains concern either license creation or license update, which together account for over 85\% of the discussions in each category. This trend reflects developers' primary concern with selecting or adjusting appropriate licenses to govern the use and distribution of their models and datasets. Notably, OSS repositories exhibit a relatively higher frequency of license update discussions (30.0\%) compared to LLMs (24.0\%), while datasets show the highest rate (43.4\%). This may suggest that dataset maintainers more frequently revise licenses to keep pace with evolving project scopes and compliance needs. In contrast, license inquiries, which are questions regarding the interpretation or usage constraints of licenses, are much more common in LLM-related discussions (21.1\%) than in OSS (3.6\%) or datasets (9.6\%). This suggests that developers working with LLMs face greater uncertainty about legal boundaries, possibly due to the complex and evolving licensing practices surrounding foundation models.
Concerns over compatibility between licenses, especially regarding model-model or model-dataset integration, are rarely discussed. Only 1.8\% of OSS issues raise license compatibility with LLMs, and 4.2\% with datasets. For LLMs and datasets themselves, compatibility discussions are almost absent, despite their practical importance in model chaining and data usage scenarios.

In summary, license selection and updating dominate the concerns in LLMware development, while compatibility and justification are less frequently addressed. This suggests an urgent need for clearer guidance and tools to support proper license selection in modern AI supply chains.

\begin{table}[t!]
  \centering
  \footnotesize
  \caption{Cumulative time span for posting and closing issues}
    \begin{tabular}{l|ccc||ccc}
    \hline
    & \multicolumn{3}{c||}{\textbf{Posting}} & \multicolumn{3}{c}{\textbf{Closing}} \\
    \cline{2-4} \cline{5-7}
    & \textbf{OSS} & \textbf{LLM} & \textbf{Dataset} & \textbf{OSS} & \textbf{LLM} & \textbf{Dataset} \\
    \hline \hline
    \textbf{0-1 Day} & 20 (6\%) & 3 (2\%) & 2 (2\%) & 204 (61\%) & 51 (30\%) & 23 (27\%) \\ \hline
    \textbf{0-7 Days} & 36 (11\%) & 24 (14\%) & 4 (5\%) & 250 (74\%) & 68 (40\%) & 35 (42\%) \\ \hline
    \textbf{0-30 Days} & 64 (19\%) & 67 (39\%) & 12 (14\%) & 274 (81\%) & 79 (46\%) & 40 (48\%) \\ \hline
    \textbf{0-90 Days} & 115 (34\%) & 99 (58\%) & 25 (30\%) & 288 (85\%) & 83 (49\%) & 43 (51\%) \\ \hline
    \textbf{0-180 Days} & 157 (47\%) & 120 (70\%) & 53 (63\%) & 292 (87\%) & 85 (50\%) & 49 (58\%) \\ \hline
    \textbf{0-1 Year} & 193 (57\%) & 140 (82\%) & 67 (80\%) & 295 (88\%) & 86 (50\%) & 50 (60\%) \\ \hline
    \textbf{0-2/10 Years} & 337 (100\%) & 171 (100\%) & 84 (100\%) & 298 (88\%) & 86 (50\%) & 50 (60\%) \\
    \hline \hline
    \textbf{Median (days)} & 310.5 & 61.2 & 120.7 & 0.1 & 0.5 & 2.4 \\ \hline
    \textbf{Mean (days)} & 645.1 & 187.6 & 252.4 & 14.3 & 12.1 & 24.1 \\
    \hline \hline
    \textbf{Still Open} & -- & -- & -- & 39 (12\%) & 85 (50\%) & 34 (40\%) \\
    \hline
    \end{tabular}
  \label{tab:time-span}
\end{table}

\smalltitle{Time Span Analysis}
To investigate how quickly license-related issues emerge, we analyze the time span from repository creation to the posting of the first license-related issue. As shown in \tabref{time-span}, the timing of issue emergence varies across artifact types. For OSS repositories, license issues arise more gradually over the project lifecycle, with 57\% posted within the first year. In contrast, LLMs and datasets see license issues emerge more rapidly after release: 82\% of LLM issues and 80\% of dataset issues are reported within the first year. This pattern may reflect the heightened scrutiny that AI artifacts receive regarding licensing, particularly given the novelty and uncertainty surrounding AI-specific licenses.
To further investigate how long license-related issues take to resolve, we analyze the duration between when an issue is opened and when it is closed. For OSS repositories, license issues are typically resolved rapidly: 61\% are closed within one day and over 80\% within the first month. By one year, nearly all issues (88\%) have been resolved, leaving only 12\% still open. In contrast, issues raised for LLMs and datasets take significantly longer to resolve. For LLMs, only 30\% are closed within one day, and approximately half (50\%) remain unresolved even after two years. Datasets exhibit a similar pattern: only 27\% are closed within a day, and 40\% remain unresolved.

Our analysis reveals a clear gap between GitHub and Hugging Face in issue resolution efficiency. While OSS projects benefit from mature processes for addressing licensing concerns, LLMs and datasets face longer delays and higher rates of unresolved issues, underscoring the need for improved license management practices and clearer guidance in the LLMware ecosystem.

\finding{License creation and update account for over 85\% of discussions in each category, suggesting that license selection remains the primary concern. LLMs and datasets show more license inquiries and slower resolution than OSS, suggesting greater uncertainty about AI-specific licenses and the need for better support.}

\subsection{RQ3: Conflict Analysis in the LLMware Supply Chain}

\subsubsection{Motivation}

License conflicts have long been recognized as a critical challenge in OSS ecosystems~\cite{li2025open,papoutsoglou2022analysis}, prompting the development of tools and approaches for automated license incompatibility detection and compliance analysis~\cite{xu2023lidetector,xu2023liresolver,liu2024catch}. More recently, the rise of LLMs has motivated studies on license usage and compatibility for AI models and datasets~\cite{wu2024large,li2025open,duan2024modelgo}. However, LLMware introduces a new form of complexity: unlike traditional OSS, it combines multiple layers of dependencies, including OSS libraries, pretrained LLMs, and datasets, each potentially governed by different license types. This multi-layered supply chain may give rise to novel conflict scenarios, such as incompatibilities between OSS licenses (e.g., \code{MIT} vs. \code{GPL}) or between OSS and AI-specific licenses (e.g., \code{OpenRAIL}, \code{LLaMA}), which have not been systematically studied. It is therefore crucial to investigate the extent to which such conflicts manifest in real-world supply chains and whether existing license analysis methods can effectively detect them.

\subsubsection{Methodology}

To detect license conflicts in LLMware supply chains, we construct a ground-truth dataset of license terms and systematically compare the licenses of upstream and downstream components. Our methodology consists of three steps:

\smalltitle{License Collection} For OSS repositories, we retrieve license files via the GitHub REST API~\cite{githubGitHubREST}, which automatically identifies licenses matching standard SPDX templates~\cite{stewart2010software,kapitsaki2017automating} and returns their license ID. Repositories with license files that deviate substantially from known templates are labeled as \code{NOASSERTION}, while those without license files are marked as \code{Not Found}.

\smalltitle{License Term Extraction} Legal terms~\cite{gangadharan2012managing,liu2019predicting} refer to clauses in a license that define the rights and obligations imposed on users. Following LiDetector~\cite{xu2023lidetector}, we represent each license as a set of predefined legal terms associated with specific attitudes: \textit{can}, \textit{cannot}, and \textit{must}. For OSS licenses, we adopt the ground truth from TLDRLegal~\cite{tldrlegal}, covering 124 common licenses. For AI-specific licenses, we manually annotated 16 representative licenses: two authors independently labeled each license against 23 terms, achieving a Cohen's $\kappa$ of 0.81, suggesting substantial inter-rater agreement. The annotation results are available in our replication package.

\newcommand{\cmark}{\text{\ding{51}}}
\newcommand{\xmark}{\text{\ding{55}}}
\newcommand{\OK}{$\cmark\xspace$}
\newcommand{\ERR}{$\xmark\xspace$}

\begin{wraptable}{r}{0.42\linewidth}
  \centering
  \scriptsize
  \caption{Compatibility between upstream and downstream components}
    \begin{tabular}{l|c|c|c}
    \hline
    \diagbox{\textbf{Down.}}{\textbf{Up.}}  & \textbf{Can} & \textbf{Cannot} & \textbf{Must} \\
    \hline \hline
    \textbf{Can} & \OK & \ERR & \ERR \\
    \hline
    \textbf{Cannot} & \OK & \OK & \ERR \\
    \hline
    \textbf{Must} & \OK & \ERR & \OK \\
    \hline
    \end{tabular}
  \tablabel{conflict-table}
\end{wraptable}

\vspace{0.1cm}
\noindent\textbf{Conflict Detection} We model the LLMware supply chain as a directed graph from downstream to upstream components, as illustrated in \figref{supply-chain}. For each adjacent pair of nodes, we compare their license term attitudes to identify potential conflicts. The key principle is that a downstream license must not be more permissive than its upstream counterpart. Terms not explicitly mentioned default to \textit{cannot} for rights and \textit{can} for obligations, following LiDetector~\cite{xu2023lidetector}.

\tabref{conflict-table} summarizes the compatibility rules between upstream and downstream licenses. Compatibility is preserved only when downstream obligations are equal to or stricter than those of the upstream. For example, if the upstream allows an action (\textit{Can}), the downstream may also allow it (\textit{Can}) or impose stricter requirements (\textit{Must}). Conversely, if the upstream prohibits an action (\textit{Cannot}), the downstream cannot override it by requiring or allowing it. Similarly, if the upstream requires an obligation (\textit{Must}), the downstream must also enforce it; otherwise, a conflict is flagged.

\begin{table*}[tp]
  \centering
  \scriptsize
  \caption{Top-10 conflicting license pairs}
  \scalebox{0.97}{
    \begin{tabular}{l|l|l||l|l|l||l|l|l}
    \hline
    \multicolumn{3}{c||}{\textbf{OSS $\mapsto$ LLM}} & \multicolumn{3}{c||}{\textbf{LLM $\mapsto$ Dataset}} & \multicolumn{3}{c}{\textbf{LLM $\mapsto$ LLM}} \\
    \hline
     \multicolumn{1}{c|}{\textbf{OSS}} & \multicolumn{1}{c|}{\textbf{LLM}} & \multicolumn{1}{c||}{\textbf{\%}} & \multicolumn{1}{c|}{\textbf{LLM}} & \multicolumn{1}{c|}{\textbf{Dataset}} & \multicolumn{1}{c||}{\textbf{\%}} & \multicolumn{1}{c|}{\textbf{LLM}} & \multicolumn{1}{c|}{\textbf{LLM}} & \multicolumn{1}{c}{\textbf{\%}} \\
    \hline\hline
    \code{No License} & \code{Apache-2.0} & 23.50\% & \code{Apache-2.0} & \code{No License} & 25.55\% & \code{Apache-2.0} & \code{MIT} & 14.29\% \\
    \hline
    \code{MIT} & \code{Apache-2.0} & 15.05\% & \code{Apache-2.0} & \code{MIT} & 14.84\% & \code{Apache-2.0} & \code{No License} & 8.16\% \\
    \hline
    \code{No License} & \code{MIT} & 14.27\% & \code{Apache-2.0} & \code{CC-BY-4.0} & 13.29\% & \code{MIT} & \code{Apache-2.0} & 8.16\% \\
    \hline
    \code{Apache-2.0} & \code{No License} & 11.68\% & \code{MIT} & \code{No License} & 5.94\% & \code{MIT} & \code{No License} & 6.12\% \\
    \hline
    \code{No License} & \code{MIT} & 10.16\% & \code{Apache-2.0} & \code{CC0-1.0} & 5.29\% & \code{No License} & \code{MIT} & 6.12\% \\
    \hline
    \code{Apache-2.0} & \code{MIT} & 7.85\% & \code{MIT} & \code{CC-BY-4.0} & 4.39\% & \code{No License} & \code{Apache-2.0} & 6.12\% \\
    \hline
    \code{No License} & \code{CC-BY-NC-4.0} & 1.34\% & \code{Apache-2.0} & \code{CC-BY-SA-4.0} & 4.26\% & \code{CC-BY-NC-4.0} & \code{Apache-2.0} & 4.08\% \\
    \hline
    \code{GPL-3.0} & \code{Apache-2.0} & 1.26\% & \code{No License} & \code{MIT} & 3.10\% & \code{Apache-2.0} & \code{LLaMA2} & 4.08\% \\
    \hline
    \code{GPL-3.0} & \code{MIT} & 0.96\% & \code{Apache-2.0} & \code{CC-BY-NC-4.0} & 2.45\% & \code{CC-BY-4.0} & \code{MIT} & 4.08\% \\
    \hline
    \code{No License} & \code{Crea. ML OpenRAIL} & 0.89\% & \code{No License} & \code{CC-BY-SA-4.0} & 2.19\% & \code{CDLA-Perm.-2.0} & \code{No License} & 4.08\% \\
    \hline
    \end{tabular}
    }
  \tablabel{top-conflicts}
\end{table*}

\subsubsection{Results}
Based on the compatibility relationships defined by legal terms and their corresponding attitudes, we analyze license conflicts along the LLMware supply chain, which we model as a directed graph. A conflict arises when two adjacent nodes (e.g., OSS $\mapsto$ LLM, LLM $\mapsto$ Dataset, or LLM $\mapsto$ LLM) contain at least one pair of terms whose attitudes violate the predefined compatibility rules. In total, we identify 68,778 supply chains, comprising 37,883 two-layer OSS $\mapsto$ LLM chains and 30,895 three-layer OSS $\mapsto$ LLM $\mapsto$ Dataset chains. Of these, 35,726 (52\%) exhibit license conflicts, consistent with prior studies~\cite{xu2023lidetector,xu2023liresolver}. \tabref{top-conflicts} summarizes the top-10 incompatible license pairs across the three types of supply-chain relationships.

\smalltitle{Conflicts between OSS and LLMs}
We observe a large number of conflicts at the OSS-to-LLM boundary, primarily driven by missing licenses and incompatibilities among traditional OSS licenses. The most frequent pair is \code{No License} $\rightarrow$ \code{Apache-2.0}, accounting for 23.50\% of all OSS $\mapsto$ LLM conflicts. This highlights a prevalent issue in LLMware: many downstream OSS repositories lack explicit licensing terms, causing upstream LLMs to inherit ambiguous or incompatible obligations. The second and third most common patterns are \code{MIT} $\rightarrow$ \code{Apache-2.0} (15.05\%) and \code{No License} $\rightarrow$ \code{MIT} (14.27\%). Interestingly, although strong-copyleft licenses (e.g., \code{GPL-3.0}) appear in the conflict list (ranks 8 and 9), their absolute frequency remains comparatively low, suggesting that LLM projects tend to avoid GPL-licensed dependencies to reduce downstream obligations.

\smalltitle{Conflicts between LLMs and Datasets}
Conflicts between models and datasets are even more pronounced, with \code{Apache-2.0} $\rightarrow$ \code{No License} being the most frequent pair (25.55\%). This reveals a systemic gap in dataset licensing practices: many datasets lack clear usage terms, making it difficult to determine whether model training violates restrictions such as attribution, share-alike, or commercial-use limitations. Pairs such as \code{Apache-2.0} $\rightarrow$ \code{CC-BY-4.0} (13.29\%) and \code{Apache-2.0} $\rightarrow$ \code{CC0-1.0} (5.29\%) highlight challenges when combining software-style permissive licenses with Creative Commons content licenses, whose attribution and share-alike terms are not designed for model-training scenarios. Conflicts involving non-commercial datasets (e.g., \code{Apache-2.0} $\rightarrow$ \code{CC-BY-NC-4.0}) also appear frequently, emphasizing that downstream model usage, especially commercial deployment, can contradict upstream dataset restrictions.

\smalltitle{Conflicts between LLMs and Base Models}
The model–base-model layer reveals a different pattern: conflicts mainly arise among permissive licenses and proprietary AI-specific licenses. The most common conflict is \code{Apache-2.0} $\rightarrow$ \code{MIT} (14.29\%), followed by \code{Apache-2.0} $\rightarrow$ \code{No License} and \code{MIT} $\rightarrow$ \code{Apache-2.0} (both 8.16\%). These conflicts are largely driven by subtle differences in attribution requirements, liability disclaimers, and patent-grant clauses when models recursively build upon previous models.

We also observe conflicts involving AI-specific licenses such as \code{LLaMA2} and \code{OpenRAIL}. For example, \code{Apache-2.0} $\rightarrow$ \code{LLaMA2} (4.08\%) and \code{No License} $\rightarrow$ \code{CreativeML-OpenRAIL-M} (0.89\%) often stem from use-case restrictions, such as limitations on harmful use or conditions on commercial use, that are incompatible with permissive OSS licenses assuming unrestricted downstream reuse. Overall, conflicts in this layer indicate that LLM developers frequently combine components governed by heterogeneous licensing philosophies, including permissive software licenses, content licenses, and AI-specific responsible-use licenses, leading to non-trivial incompatibilities along the model inheritance chain.

\finding{License conflicts are prevalent in LLMware supply chains, with 52\% exhibiting at least one conflict. Unlike traditional OSS, a substantial portion of these conflicts involves AI-specific licenses, highlighting a new class of compatibility challenges unique to LLMware ecosystems.}

\subsection{RQ4: Evaluation of License Analysis Approaches on LLMware}

\subsubsection{Motivation}
The results of RQ3 demonstrate that license conflicts are pervasive in LLMware supply chains, spanning OSS repositories, pretrained models, and datasets. While this reveals the severity of the problem, it does not indicate whether existing license compatibility analysis approaches can effectively address such conflicts in practice. Most existing approaches were designed primarily for traditional OSS licenses, whereas AI-specific licenses, which are common in LLMware supply chains, are largely overlooked or unsupported. In this RQ, we evaluate representative license compatibility analysis approaches on their ability to detect conflicts within LLMware, assessing their effectiveness and identifying limitations in this new context. Motivated by recent advances in LLMs and LLM-based agents~\cite{liu2024large,he2025llm,hu2025assessing,terragni2025future}, we further propose \textsc{LiAgent}, an LLM-based agent approach that leverages the reasoning capability of LLMs to analyze license terms and detect potential incompatibilities, and evaluate its effectiveness against existing methods.

\subsubsection{Methodology}
This subsection describes the dataset, baselines, and our approach.

\smalltitle{Dataset}
To support the evaluation, we construct a benchmark dataset comprising four sets: traditional OSS licenses (\textbf{OSS}), AI-specific licenses (\textbf{AI}), mutated OSS licenses (\textbf{OSS-Mut}), and mutated AI-specific licenses (\textbf{AI-Mut}). Each license instance is associated with a set of legal terms, their attitudes (\textit{can}, \textit{cannot}, \textit{must}), and supporting sentence(s) from the original license text. The mutated licenses are generated to enlarge the dataset by selecting one legal term per license and changing its attitude to each of the other two values among \textit{can}, \textit{cannot}, and \textit{must}, while keeping all other terms unchanged. If a term appears in multiple sentences, all corresponding sentences are mutated consistently. Each mutation modifies exactly one term.

For the 16 AI-specific licenses, two authors independently annotated the attitudes of 23 legal terms following the setup of the previous RQ, with a third author resolving disagreements to establish the final ground truth. Cohen's $\kappa$ between annotators is 0.81, indicating substantial agreement. For each term–attitude pair, we also record the exact sentence(s) that justify the annotation. The annotation results are available in our replication package. The resulting four license sets are:
\begin{itemize}[leftmargin=*]
    \item \textbf{OSS}: 124 commonly used traditional OSS licenses from TLDRLegal, which provides ground-truth mappings between licenses and legal terms~\cite{li2023lisum,xu2023lidetector}.
    \item \textbf{OSS-Mut}: 620 mutated versions of the traditional OSS licenses.
    \item \textbf{AI}: 16 representative AI-specific licenses collected from Hugging Face model cards, excluding standard OSS licenses.
    \item \textbf{AI-Mut}: 176 mutated versions of the AI-specific licenses.
\end{itemize}

\smalltitle{License Analysis Approaches} We include LiDetector~\cite{xu2023lidetector}, the state-of-the-art license analysis approach for OSS supply chains, in our evaluation. LiDetector also identified a strong baseline that uses semantic similarity matching~\cite{karvelis2018topic} for term extraction combined with SST-based sentiment analysis~\cite{socher2013recursive} for attitude identification. We evaluate this baseline as well, referred to as \textit{Semantic Similarity + SST-based Sentiment Analysis}.


\smalltitle{Our Approach: \ourtool} LLM agents have demonstrated strong capabilities in code and text analysis tasks~\cite{liu2024large,he2025llm,hu2025assessing,terragni2025future}. Motivated by these advances, we propose \ourtool, an LLM-based multi-agent framework for license term extraction and compatibility analysis. \ourtool consists of two agents: an \textit{extraction agent} and a \textit{repair agent}. The extraction agent takes the full license text as input and identifies legal terms along with their corresponding attitudes (\textit{can}, \textit{cannot}, \textit{must}), as well as the supporting sentence(s) for each identified term. As shown in \figref{term-extraction}, we design a structured prompt that includes: (1) task instructions defining the extraction goal, (2) a predefined list of 23 legal terms with descriptions to guide the agent, and (3) few-shot examples demonstrating the expected JSON output format. After extraction, we perform an \textit{attitude consistency check}: if the same term is assigned conflicting attitudes across different sentences (e.g., both \textit{can} and \textit{cannot}), the \textit{repair agent} is invoked. The repair agent takes the conflicting extractions and the original license text as input, re-analyzes the context, and produces a revised, consistent assignment. This process iterates for up to three rounds until consistency is achieved or the iteration limit is reached; we set this limit based on preliminary experiments showing that most inconsistencies are resolved within two iterations. 

Since prior studies have shown that different backbone models can yield varying performance~\cite{he2025llm,hu2025assessing}, we evaluate \ourtool with four mainstream LLMs: GPT-4o, GPT-4o-Mini, DeepSeek-V3-671b~\cite{liu2024deepseek}, and Qwen-Max~\cite{yang2025qwen3}, to assess the generalizability of our approach.

\begin{figure}[t]
\begin{tcolorbox}[tile, size=fbox,left=2mm, right=2mm, boxrule=0pt, top=1mm, bottom=1mm, colback=blue!3!white, 
    sharp corners=south]
\scriptsize
\textbf{[Instruction]}: You are a legal analysis assistant specializing in license agreements. Your task is to analyze the given text and identify key legal terms, categorize actions, and infer rules or conditions for these actions. Always respond in a structured and concise format. Use JSON for output as specified. Ensure that the output strictly follows the format provided in the example. Analyze the following license text and complete the tasks below:

\vspace{0.05cm}
\textbf{[The License Content]}: \textbf{\{LICENSE\}}

\vspace{0.05cm}
\textbf{[Instruction]}: The following is a list of key legal terms and their meanings. For each term, only consider the provided interpretation when analyzing the license text...

\vspace{0.05cm}
\textbf{[Term Descriptions]}: \{`Distribute': `Describes the ability to distribute original or modified (derivative) works.', `Commercial Use': ...\}

\vspace{0.05cm}
\textbf{[Example Input]}: ``You are allowed to distribute modified works. You must give credit to the original author of the work. You can modify if there are state changes.'' 

\vspace{0.05cm}
\textbf{[Example Output]}:  [\{"sentence": "You are allowed to distribute modified works", "term": "Distribute", "attitude": "can", ...\}, ...]

\end{tcolorbox}
\caption{The prompt template of \ourtool for term extraction}
\figlabel{term-extraction}
\end{figure}
\begin{table}[tp]
  \centering
  \scriptsize
  \caption{The license analysis results}
    \begin{tabular}{l|c|c|c || c|c|c || c|c|c || c|c|c}
    \hline
    \multirow{2}{*}{} & \multicolumn{3}{c||}{\textbf{OSS (124)}} & \multicolumn{3}{c||}{\textbf{OSS-Mut (620)}} & \multicolumn{3}{c||}{\textbf{AI (16)}} & \multicolumn{3}{c}{\textbf{AI-Mut (176)}} \\
\cline{2-13}       & \textbf{P.} & \textbf{R.} & \textbf{F1} & \textbf{P.} & \textbf{R.} & \textbf{F1} & \textbf{P.} & \textbf{R.} & \textbf{F1} & \textbf{P.} & \textbf{R.} & \textbf{F1} \\
    \hline \hline
    \textbf{Seman. Sim. + Senti. Ana.} & 56\% & 68\% & 61\% & 57\% & 67\% & 61\% & 34\% & 56\% & 42\% & 33\% & 55\% & 41\% \\
    \hline
    \textbf{LiDetector} & 71\% & 81\% & 76\% & 71\% & 80\% & 75\% & 81\% & 80\% & 81\% & 80\% & 78\% & 79\% \\
    \hline \hline
    \textbf{\ourtool (GPT-4o)} & \cellcolor{gray!40}91\% & 83\% & 87\% & \cellcolor{gray!40}89\% & 82\% & 85\% & \cellcolor{gray!40}93\% & 82\% & 87\% & \cellcolor{gray!40}93\% & 83\% & \cellcolor{gray!40}88\% \\
    \hline
    \textbf{\ourtool (GPT-4o-Mini)} & 88\% & 82\% & 85\% & 86\% & 81\% & 83\% & 86\% & 82\% & 84\% & 87\% & 79\% & 83\% \\
    \hline
    \textbf{\ourtool (DeepSeek-V3)} & 87\% & \cellcolor{gray!40}88\% & \cellcolor{gray!40}88\% & 86\% & \cellcolor{gray!40}86\% & \cellcolor{gray!40}86\% & 90\% & \cellcolor{gray!40}89\% & \cellcolor{gray!40}89\% & 88\% & \cellcolor{gray!40}88\% & \cellcolor{gray!40}88\% \\
    \hline
    \textbf{\ourtool (Qwen-Max)} & 88\% & 84\% & 86\% & 87\% & 83\% & 85\% & 89\% & 86\% & 87\% & 87\% & 84\% & 86\% \\
    \hline
    \end{tabular}
  \tablabel{rq4-res}
\end{table}

\subsubsection{Results}

\tabref{rq4-res} reports the performance of different license analysis approaches on the four benchmark collections.

Overall, we observe clear performance differences among the evaluated approaches. The \textit{Semantic Similarity + SST-based Sentiment Analysis} baseline achieves 61\% F1 on OSS licenses but only 42\% on AI-specific licenses, indicating limited generalizability to AI-specific licensing terms. LiDetector, the state-of-the-art approach, performs substantially better, achieving 76\% F1 on OSS and 81\% on AI. In contrast, all variants of \ourtool consistently outperform both baselines. Among the four backbone LLMs, DeepSeek-V3 achieves the best overall performance with 88\% F1 on OSS and 89\% on AI, outperforming LiDetector by up to 12\% and 8\%, respectively. GPT-4o and Qwen-Max also demonstrate strong results, achieving 87\% and 86\% F1 on OSS, respectively, while GPT-4o-Mini shows slightly lower but still competitive performance (85\% on OSS, 84\% on AI).

On mutated licenses, all approaches exhibit performance degradation, indicating that attitude perturbations pose additional challenges for license analysis. The baseline maintains 61\% F1 on OSS-Mut but drops to 41\% on AI-Mut, suggesting high sensitivity to wording variations in AI-specific licenses. LiDetector shows more stability, decreasing only slightly to 75\% on OSS-Mut and 79\% on AI-Mut. In contrast, \ourtool demonstrates the strongest robustness across all variants. DeepSeek-V3 achieves 86\% F1 on OSS-Mut and 88\% on AI-Mut, with reductions of no more than 2\% compared to the original collections. Similarly, GPT-4o maintains 85\% on OSS-Mut and 88\% on AI-Mut. These results suggest that LLM-based agents better accommodate the heterogeneous language patterns and attitude perturbations commonly found in real-world license modifications, making them more suitable for analyzing the diverse licensing landscape in LLMware ecosystems.

\finding{\ourtool consistently outperforms existing approaches across all four benchmark collections, achieving up to 89\% F1 compared to 81\% for the state-of-the-art. Results suggest that LLM-based agents better handle heterogeneous language patterns in licenses, making them more suitable for LLMware's diverse licensing landscape.}

\section{Discussion}

\begin{table}[tp!]
  \centering
  \scriptsize
  \caption{Status of the reported license issues to LLMware systems}
    \begin{tabular}{l|l|l|l|l|c}
    \hline
    \textbf{ID} & \textbf{Type} & \textbf{Downstream License} & \textbf{Upstream License} & \textbf{Repo Name} & \textbf{Fixed} \\
    \hline \hline
    \textbf{1} & OSS $\mapsto$ LLM & \code{MIT} &  \code{CreativeML Open RAIL-M} & xxx/erasing & \OK \\
    \hline
    \textbf{2} & OSS $\mapsto$ LLM & \code{MIT} &  \code{CreativeML Open RAIL-M} & xxx/MaterialPalette & \OK \\
    \hline
    \textbf{3} & OSS $\mapsto$ LLM & \code{MIT} &  \code{OpenRAIL++} & xxx/sliders & \OK \\
    \hline
    \textbf{4} & OSS $\mapsto$ LLM & \code{MIT} &  \code{OpenRAIL++} & xxx/CRM & \ERR \\
    \hline
    \textbf{5} & OSS $\mapsto$ LLM & \code{MIT} & \code{OpenRAIL++} & xxx/realfill & \OK \\
    \hline
    \textbf{6} & OSS $\mapsto$ LLM & \code{MIT} &  \code{LLaMA2} & xxx/HELMET & \ERR \\
    \hline
    \textbf{7} & OSS $\mapsto$ LLM & \code{MIT} & \code{CC-BY-4.0} & xxx/AIlice & \OK \\
    \hline
    \textbf{8} & OSS $\mapsto$ LLM & \code{Apache-2.0} & \code{CC-BY-NC-4.0} & xxx/CleanS2S & \OK \\
    \hline
    \textbf{9} & OSS $\mapsto$ LLM & \code{Apache-2.0} & \code{CC-BY-NC-4.0} & xxx/dingo & \OK \\
    \hline
    \textbf{10} & LLM $\mapsto$ Dataset & \code{Apache-2.0} & \code{CC-BY-4.0} & xxx/text2vec-base-chinese & \ERR \\
    \hline
    \textbf{11} & LLM $\mapsto$ Dataset & \code{Apache-2.0} & \code{CC-BY-NC-4.0} & xxx/deberta-v3-base-prompt-injection & \OK \\
    \hline
    \end{tabular}
  \tablabel{reports}
\end{table}

\subsection{Practical Application of \ourtool in Real-world LLMware Systems}
To assess the practical usefulness of \ourtool, we collected feedback from LLMware developers on the license incompatibility issues detected by our approach. Since reporting all detected issues is impractical, we systematically selected and reported 60 issues to real-world LLMware projects, prioritizing those with higher GitHub stars or Hugging Face likes. The selection was based on three criteria: (1) target projects are active and widely used; (2) incompatibilities are actionable, i.e., resolvable by modifying or clarifying license declarations; and (3) issues cover a diverse set of OSS and AI-specific licenses.

At the time of writing, 11 of 60 reported issues have been confirmed by developers, 9 were denied, and the remaining 40 await responses. Among the confirmed issues, 8 have already been fixed, as shown in \tabref{reports}. Notably, 6 of the 11 confirmed issues involve AI-specific licenses such as \textsc{LLaMA2}, \textsc{OpenRAIL++}, and \textsc{CreativeML Open RAIL-M}, highlighting the practical importance of AI-specific license compatibility analysis. Moreover, two of the confirmed issues are associated with highly influential LLMs with over 107 million and 5 million downloads on Hugging Face, respectively, suggesting that such incompatibilities may impact a large number of downstream projects and users. These results demonstrate that \ourtool can effectively uncover practical and previously unnoticed license issues in real-world LLMware systems.

\subsection{Lessons Learned}
We derive several lessons regarding license incompatibility in LLMware from our empirical study and developer feedback.

\smalltitle{Lesson 1: Heterogeneous licensing paradigms are a primary source of conflicts} Incompatibilities often stem from mixing traditional OSS licenses with AI-specific licenses that impose usage-based restrictions. Many developers assume permissive licenses (e.g., \textsc{MIT}, \textsc{Apache-2.0}) are universally compatible, overlooking that AI-specific licenses (e.g., \textsc{LLaMA2}, \textsc{OpenRAIL++}) may limit commercial use or downstream deployment.

\smalltitle{Lesson 2: Resolving conflicts is often impractical} Unlike traditional OSS where replacing a module or library can address license conflicts, substituting a model or dataset is often infeasible, as they may be tightly coupled with training pipelines and downstream applications. Moreover, some incompatibilities arise from inherent conflicts between software-oriented and content-oriented licenses that cannot be resolved through simple replacement.

\smalltitle{Lesson 3: Respect the hierarchical nature of supply chains} Downstream components should not adopt licenses more permissive than their upstream dependencies or more restrictive than higher-level modules containing them. Understanding this hierarchy is essential for avoiding incompatibilities in multi-layered LLMware supply chains.

\subsection{Threats to Validity}
\smalltitle{Internal Validity} The results of this study rely on several manually constructed and annotated datasets, including the term–attitude ground truth for OSS and AI-specific licenses and the labeled developer discussions. Although multiple authors independently performed the annotations and we measured substantial inter-rater agreement, labeling errors or subjective interpretations may still exist. To mitigate this threat, disagreements were resolved through discussion and cross-validation, and a senior researcher was involved in the final decision-making process.

\smalltitle{External Validity} Our supply-chain construction relies on Sourcegraph over repositories with at least 5 GitHub stars, which may under-represent less popular projects. However, this threshold is commonly adopted in software engineering research~\cite{niu2026trustinsecuredemystifyingdevelopers,DBLP:journals/ese/MunaiahKCN17,DBLP:journals/jss/BorgesV18} to ensure code quality and reduce noise. Additionally, we focus on GitHub and Hugging Face, the two most prominent platforms that cover the majority of LLMware supply chains in practice.

\smalltitle{Construct Validity} We characterize licenses using 23 legal terms and three attitudes, which may not fully capture nuanced clauses in AI-specific licenses. However, our term set is derived from established legal frameworks SPDX~\cite{spdx} and covers the most common licensing obligations. Moreover, our conflict detection relies on predefined compatibility rules adopted from LiDetector~\cite{xu2023lidetector}, which have been validated in prior studies and align with widely accepted legal interpretations.
\section{Conclusion and Future Work}
This paper presents the first large-scale empirical study of licensing practices and incompatibilities in LLMware ecosystems. By analyzing 12,180 GitHub repositories, 3,988 LLMs, and 708 datasets, we reveal that license conflicts are prevalent, affecting 52\% of supply chains, and exhibit patterns distinct from traditional OSS settings, particularly involving AI-specific licenses such as LLaMA and OpenRAIL. Our analysis of developer discussions further highlights that license selection and maintenance remain primary pain points, with LLMs and datasets facing slower issue resolution than OSS projects. We also find that existing approaches perform poorly on AI-specific licenses, motivating us to propose \ourtool, an LLM-based multi-agent framework that achieves up to 89\% F1 and outperforms the state-of-the-art by 12 percentage points. We reported 60 detected issues to real-world projects, with 11 confirmed and 8 fixed. Notably, two affected LLMs have over 107 million and 5 million downloads on Hugging Face, respectively, suggesting that such issues may impact a large number of downstream applications. For future work, we plan to extend \ourtool with automated license recommendation and conflict resolution capabilities and explore its application to other AI artifact platforms beyond Hugging Face.


\section*{Data Availability}
For reproducibility and to advance the state of research, we have made our curated dataset and code anonymous and publicly available at \url{https://anonymous.4open.science/r/LiAgent-DB35/}.

\bibliographystyle{ACM-Reference-Format}
\bibliography{ref}


\begin{thebibliography}{79}


\ifx \showCODEN    \undefined \def \showCODEN     #1{\unskip}     \fi
\ifx \showISBNx    \undefined \def \showISBNx     #1{\unskip}     \fi
\ifx \showISBNxiii \undefined \def \showISBNxiii  #1{\unskip}     \fi
\ifx \showISSN     \undefined \def \showISSN      #1{\unskip}     \fi
\ifx \showLCCN     \undefined \def \showLCCN      #1{\unskip}     \fi
\ifx \shownote     \undefined \def \shownote      #1{#1}          \fi
\ifx \showarticletitle \undefined \def \showarticletitle #1{#1}   \fi
\ifx \showURL      \undefined \def \showURL       {\relax}        \fi
\providecommand\bibfield[2]{#2}
\providecommand\bibinfo[2]{#2}
\providecommand\natexlab[1]{#1}
\providecommand\showeprint[2][]{arXiv:#2}

\bibitem[Borges and Valente(2018)]%
        {DBLP:journals/jss/BorgesV18}
\bibfield{author}{\bibinfo{person}{Hudson Borges} {and} \bibinfo{person}{Marco~T{\'{u}}lio Valente}.} \bibinfo{year}{2018}\natexlab{}.
\newblock \showarticletitle{What's in a GitHub Star? Understanding Repository Starring Practices in a Social Coding Platform}.
\newblock \bibinfo{journal}{\emph{J. Syst. Softw.}}  \bibinfo{volume}{146} (\bibinfo{year}{2018}), \bibinfo{pages}{112--129}.
\newblock
\href{https://doi.org/10.1016/J.JSS.2018.09.016}{doi:\nolinkurl{10.1016/J.JSS.2018.09.016}}


\bibitem[Chang et~al\mbox{.}(2024)]%
        {10.1145/3641289}
\bibfield{author}{\bibinfo{person}{Yupeng Chang}, \bibinfo{person}{Xu Wang}, \bibinfo{person}{Jindong Wang}, \bibinfo{person}{Yuan Wu}, \bibinfo{person}{Linyi Yang}, \bibinfo{person}{Kaijie Zhu}, \bibinfo{person}{Hao Chen}, \bibinfo{person}{Xiaoyuan Yi}, \bibinfo{person}{Cunxiang Wang}, \bibinfo{person}{Yidong Wang}, \bibinfo{person}{Wei Ye}, \bibinfo{person}{Yue Zhang}, \bibinfo{person}{Yi Chang}, \bibinfo{person}{Philip~S. Yu}, \bibinfo{person}{Qiang Yang}, {and} \bibinfo{person}{Xing Xie}.} \bibinfo{year}{2024}\natexlab{}.
\newblock \showarticletitle{A Survey on Evaluation of Large Language Models}.
\newblock \bibinfo{journal}{\emph{ACM Trans. Intell. Syst. Technol.}} \bibinfo{volume}{15}, \bibinfo{number}{3}, Article \bibinfo{articleno}{39} (\bibinfo{date}{March} \bibinfo{year}{2024}), \bibinfo{numpages}{45}~pages.
\newblock
\showISSN{2157-6904}
\href{https://doi.org/10.1145/3641289}{doi:\nolinkurl{10.1145/3641289}}


\bibitem[Chen et~al\mbox{.}(2025)]%
        {chen2025deep}
\bibfield{author}{\bibinfo{person}{Xiangping Chen}, \bibinfo{person}{Xing Hu}, \bibinfo{person}{Yuan Huang}, \bibinfo{person}{He Jiang}, \bibinfo{person}{Weixing Ji}, \bibinfo{person}{Yanjie Jiang}, \bibinfo{person}{Yanyan Jiang}, \bibinfo{person}{Bo Liu}, \bibinfo{person}{Hui Liu}, \bibinfo{person}{Xiaochen Li}, {et~al\mbox{.}}} \bibinfo{year}{2025}\natexlab{}.
\newblock \showarticletitle{Deep Learning-based Software Engineering: Progress, Challenges, and Opportunities}.
\newblock \bibinfo{journal}{\emph{Science China Information Sciences}} \bibinfo{volume}{68}, \bibinfo{number}{1} (\bibinfo{year}{2025}), \bibinfo{pages}{1--88}.
\newblock


\bibitem[Cox(2025)]%
        {10.1145/3762635}
\bibfield{author}{\bibinfo{person}{Russ Cox}.} \bibinfo{year}{2025}\natexlab{}.
\newblock \showarticletitle{Fifty Years of Open Source Software Supply-Chain Security}.
\newblock \bibinfo{journal}{\emph{Commun. ACM}} \bibinfo{volume}{68}, \bibinfo{number}{10} (\bibinfo{date}{Sept.} \bibinfo{year}{2025}), \bibinfo{pages}{88–95}.
\newblock
\showISSN{0001-0782}
\href{https://doi.org/10.1145/3762635}{doi:\nolinkurl{10.1145/3762635}}


\bibitem[Cram{\'e}r(1946)]%
        {cramér1946mathematical}
\bibfield{author}{\bibinfo{person}{H. Cram{\'e}r}.} \bibinfo{year}{1946}\natexlab{}.
\newblock \bibinfo{booktitle}{\emph{Mathematical Methods of Statistics}}.
\newblock \bibinfo{publisher}{Princeton University Press}.
\newblock
\showISBNx{9780691080048}
\showLCCN{a46005922}
\urldef\tempurl%
\url{https://books.google.com.sg/books?id=_db1jwEACAAJ}
\showURL{%
\tempurl}


\bibitem[Cui et~al\mbox{.}(2025)]%
        {cui2025exploring}
\bibfield{author}{\bibinfo{person}{Xing Cui}, \bibinfo{person}{Jingzheng Wu}, \bibinfo{person}{Xiang Ling}, \bibinfo{person}{Tianyue Luo}, \bibinfo{person}{Mutian Yang}, {and} \bibinfo{person}{Wenxiang Ou}.} \bibinfo{year}{2025}\natexlab{}.
\newblock \showarticletitle{Exploring Large Language Models for Analyzing Open Source License Conflicts: How Far Are We?}. In \bibinfo{booktitle}{\emph{2025 IEEE/ACM 47th International Conference on Software Engineering: Companion Proceedings (ICSE-Companion)}}. \bibinfo{pages}{291--302}.
\newblock


\bibitem[Cui et~al\mbox{.}(2023)]%
        {cui2023empirical}
\bibfield{author}{\bibinfo{person}{Xing Cui}, \bibinfo{person}{Jingzheng Wu}, \bibinfo{person}{Yanjun Wu}, \bibinfo{person}{Xu Wang}, \bibinfo{person}{Tianyue Luo}, \bibinfo{person}{Sheng Qu}, \bibinfo{person}{Xiang Ling}, {and} \bibinfo{person}{Mutian Yang}.} \bibinfo{year}{2023}\natexlab{}.
\newblock \showarticletitle{An empirical study of license conflict in free and open source software}. In \bibinfo{booktitle}{\emph{2023 IEEE/ACM 45th International Conference on Software Engineering: Software Engineering in Practice (ICSE-SEIP)}}. IEEE, \bibinfo{pages}{495--505}.
\newblock


\bibitem[Duan et~al\mbox{.}(2024)]%
        {duan2024modelgo}
\bibfield{author}{\bibinfo{person}{Moming Duan}, \bibinfo{person}{Qinbin Li}, {and} \bibinfo{person}{Bingsheng He}.} \bibinfo{year}{2024}\natexlab{}.
\newblock \showarticletitle{Modelgo: A practical tool for machine learning license analysis}. In \bibinfo{booktitle}{\emph{Proceedings of the ACM Web Conference 2024}}. \bibinfo{pages}{1158--1169}.
\newblock


\bibitem[Face(2026)]%
        {huggingfaceModelCards}
\bibfield{author}{\bibinfo{person}{Hugging Face}.} \bibinfo{year}{2026}\natexlab{}.
\newblock \bibinfo{title}{{M}odel {C}ards}.
\newblock \bibinfo{howpublished}{\url{https://huggingface.co/docs/hub/model-cards\#model-card-metadata}}.
\newblock
\newblock
\shownote{[Accessed 30-01-2026]}.


\bibitem[Fan et~al\mbox{.}(2023)]%
        {10449667}
\bibfield{author}{\bibinfo{person}{Angela Fan}, \bibinfo{person}{Beliz Gokkaya}, \bibinfo{person}{Mark Harman}, \bibinfo{person}{Mitya Lyubarskiy}, \bibinfo{person}{Shubho Sengupta}, \bibinfo{person}{Shin Yoo}, {and} \bibinfo{person}{Jie~M. Zhang}.} \bibinfo{year}{2023}\natexlab{}.
\newblock \showarticletitle{Large Language Models for Software Engineering: Survey and Open Problems}. In \bibinfo{booktitle}{\emph{2023 IEEE/ACM International Conference on Software Engineering: Future of Software Engineering (ICSE-FoSE)}}. \bibinfo{pages}{31--53}.
\newblock
\href{https://doi.org/10.1109/ICSE-FoSE59343.2023.00008}{doi:\nolinkurl{10.1109/ICSE-FoSE59343.2023.00008}}


\bibitem[Franke et~al\mbox{.}(2012)]%
        {doi:10.1177/1098214011426594}
\bibfield{author}{\bibinfo{person}{Todd~Michael Franke}, \bibinfo{person}{Timothy Ho}, {and} \bibinfo{person}{Christina~A. Christie}.} \bibinfo{year}{2012}\natexlab{}.
\newblock \showarticletitle{The Chi-Square Test: Often Used and More Often Misinterpreted}.
\newblock \bibinfo{journal}{\emph{American Journal of Evaluation}} \bibinfo{volume}{33}, \bibinfo{number}{3} (\bibinfo{year}{2012}), \bibinfo{pages}{448--458}.
\newblock
\showeprint{https://doi.org/10.1177/1098214011426594}
\href{https://doi.org/10.1177/1098214011426594}{doi:\nolinkurl{10.1177/1098214011426594}}


\bibitem[Fu et~al\mbox{.}(2024)]%
        {10495699}
\bibfield{author}{\bibinfo{person}{Daocheng Fu}, \bibinfo{person}{Xin Li}, \bibinfo{person}{Licheng Wen}, \bibinfo{person}{Min Dou}, \bibinfo{person}{Pinlong Cai}, \bibinfo{person}{Botian Shi}, {and} \bibinfo{person}{Yu Qiao}.} \bibinfo{year}{2024}\natexlab{}.
\newblock \showarticletitle{Drive Like a Human: Rethinking Autonomous Driving with Large Language Models}. In \bibinfo{booktitle}{\emph{2024 IEEE/CVF Winter Conference on Applications of Computer Vision Workshops (WACVW)}}. \bibinfo{pages}{910--919}.
\newblock
\href{https://doi.org/10.1109/WACVW60836.2024.00102}{doi:\nolinkurl{10.1109/WACVW60836.2024.00102}}


\bibitem[Gangadharan et~al\mbox{.}(2012)]%
        {gangadharan2012managing}
\bibfield{author}{\bibinfo{person}{GR Gangadharan}, \bibinfo{person}{Vincenzo D’Andrea}, \bibinfo{person}{Stefano De~Paoli}, {and} \bibinfo{person}{Michael Weiss}.} \bibinfo{year}{2012}\natexlab{}.
\newblock \showarticletitle{Managing license compliance in free and open source software development}.
\newblock \bibinfo{journal}{\emph{Information Systems Frontiers}} \bibinfo{volume}{14}, \bibinfo{number}{2} (\bibinfo{year}{2012}), \bibinfo{pages}{143--154}.
\newblock


\bibitem[German et~al\mbox{.}(2010)]%
        {german2010sentence}
\bibfield{author}{\bibinfo{person}{Daniel~M German}, \bibinfo{person}{Yuki Manabe}, {and} \bibinfo{person}{Katsuro Inoue}.} \bibinfo{year}{2010}\natexlab{}.
\newblock \showarticletitle{A sentence-matching method for automatic license identification of source code files}. In \bibinfo{booktitle}{\emph{Proceedings of the 25th IEEE/ACM International Conference on Automated Software Engineering}}. \bibinfo{pages}{437--446}.
\newblock


\bibitem[{G}it{H}ub(2026a)]%
        {copilot}
\bibfield{author}{\bibinfo{person}{{G}it{H}ub}.} \bibinfo{year}{2026}\natexlab{a}.
\newblock \bibinfo{title}{{G}it{H}ub {C}opilot · {Y}our {A}{I} pair programmer}.
\newblock \bibinfo{howpublished}{\url{https://github.com/features/copilot}}.
\newblock
\newblock
\shownote{[Accessed 28-01-2026]}.


\bibitem[{G}it{H}ub(2026b)]%
        {githubGitHubREST}
\bibfield{author}{\bibinfo{person}{{G}it{H}ub}.} \bibinfo{year}{2026}\natexlab{b}.
\newblock \bibinfo{title}{{G}it{H}ub {R}{E}{S}{T} {A}{P}{I} documentation | {G}it{H}ub {D}ocs}.
\newblock \bibinfo{howpublished}{\url{https://docs.github.com/en/rest}}.
\newblock
\newblock
\shownote{[Accessed 30-01-2026]}.


\bibitem[Gobeille(2008)]%
        {gobeille2008fossology}
\bibfield{author}{\bibinfo{person}{Robert Gobeille}.} \bibinfo{year}{2008}\natexlab{}.
\newblock \showarticletitle{The fossology project}. In \bibinfo{booktitle}{\emph{Proceedings of the 2008 international working conference on Mining software repositories}}. \bibinfo{pages}{47--50}.
\newblock


\bibitem[Han et~al\mbox{.}(2026)]%
        {han2026cloud}
\bibfield{author}{\bibinfo{person}{Fanyu Han}, \bibinfo{person}{Shengyu Zhao}, \bibinfo{person}{Xiaoya Xia}, {and} \bibinfo{person}{Wei Wang}.} \bibinfo{year}{2026}\natexlab{}.
\newblock \showarticletitle{Are cloud providers exploiting open-source? An exploratory study of Redis license change}.
\newblock \bibinfo{journal}{\emph{Information and Software Technology}}  \bibinfo{volume}{190} (\bibinfo{year}{2026}), \bibinfo{pages}{107951}.
\newblock


\bibitem[He et~al\mbox{.}(2025)]%
        {he2025llm}
\bibfield{author}{\bibinfo{person}{Junda He}, \bibinfo{person}{Christoph Treude}, {and} \bibinfo{person}{David Lo}.} \bibinfo{year}{2025}\natexlab{}.
\newblock \showarticletitle{LLM-Based Multi-Agent Systems for Software Engineering: Literature Review, Vision, and the Road Ahead}.
\newblock \bibinfo{journal}{\emph{ACM Trans. Softw. Eng. Methodol.}} \bibinfo{volume}{34}, \bibinfo{number}{5}, Article \bibinfo{articleno}{124} (\bibinfo{date}{May} \bibinfo{year}{2025}), \bibinfo{numpages}{30}~pages.
\newblock


\bibitem[Hou et~al\mbox{.}(2024)]%
        {10.1145/3695988}
\bibfield{author}{\bibinfo{person}{Xinyi Hou}, \bibinfo{person}{Yanjie Zhao}, \bibinfo{person}{Yue Liu}, \bibinfo{person}{Zhou Yang}, \bibinfo{person}{Kailong Wang}, \bibinfo{person}{Li Li}, \bibinfo{person}{Xiapu Luo}, \bibinfo{person}{David Lo}, \bibinfo{person}{John Grundy}, {and} \bibinfo{person}{Haoyu Wang}.} \bibinfo{year}{2024}\natexlab{}.
\newblock \showarticletitle{Large Language Models for Software Engineering: A Systematic Literature Review}.
\newblock \bibinfo{journal}{\emph{ACM Trans. Softw. Eng. Methodol.}} \bibinfo{volume}{33}, \bibinfo{number}{8}, Article \bibinfo{articleno}{220} (\bibinfo{date}{Dec.} \bibinfo{year}{2024}), \bibinfo{numpages}{79}~pages.
\newblock
\showISSN{1049-331X}
\href{https://doi.org/10.1145/3695988}{doi:\nolinkurl{10.1145/3695988}}


\bibitem[Hu et~al\mbox{.}(2025b)]%
        {10.1145/3713081.3731747}
\bibfield{author}{\bibinfo{person}{Qiang Hu}, \bibinfo{person}{Xiaofei Xie}, \bibinfo{person}{Sen Chen}, \bibinfo{person}{Lili Quan}, {and} \bibinfo{person}{Lei Ma}.} \bibinfo{year}{2025}\natexlab{b}.
\newblock \showarticletitle{Large Language Model Supply Chain: Open Problems From the Security Perspective}. In \bibinfo{booktitle}{\emph{Proceedings of the 34th ACM SIGSOFT International Symposium on Software Testing and Analysis}} (Clarion Hotel Trondheim, Trondheim, Norway) \emph{(\bibinfo{series}{ISSTA Companion '25})}. \bibinfo{publisher}{Association for Computing Machinery}, \bibinfo{address}{New York, NY, USA}, \bibinfo{pages}{169–173}.
\newblock
\showISBNx{9798400714740}
\href{https://doi.org/10.1145/3713081.3731747}{doi:\nolinkurl{10.1145/3713081.3731747}}


\bibitem[Hu et~al\mbox{.}(2025a)]%
        {hu2025assessing}
\bibfield{author}{\bibinfo{person}{Xing Hu}, \bibinfo{person}{Feifei Niu}, \bibinfo{person}{Junkai Chen}, \bibinfo{person}{Xin Zhou}, \bibinfo{person}{Junwei Zhang}, \bibinfo{person}{Junda He}, \bibinfo{person}{Xin Xia}, {and} \bibinfo{person}{David Lo}.} \bibinfo{year}{2025}\natexlab{a}.
\newblock \showarticletitle{Assessing and advancing benchmarks for evaluating large language models in software engineering tasks}.
\newblock \bibinfo{journal}{\emph{ACM Transactions on Software Engineering and Methodology}} (\bibinfo{year}{2025}).
\newblock


\bibitem[Huang et~al\mbox{.}(2024)]%
        {huang2024your}
\bibfield{author}{\bibinfo{person}{Kaifeng Huang}, \bibinfo{person}{Yingfeng Xia}, \bibinfo{person}{Bihuan Chen}, \bibinfo{person}{Siyang He}, \bibinfo{person}{Huazheng Zeng}, \bibinfo{person}{Zhuotong Zhou}, \bibinfo{person}{Jin Guo}, {and} \bibinfo{person}{Xin Peng}.} \bibinfo{year}{2024}\natexlab{}.
\newblock \showarticletitle{Your “Notice” Is Missing: Detecting and Fixing Violations of Modification Terms in Open Source Licenses during Forking}. In \bibinfo{booktitle}{\emph{Proceedings of the 33rd ACM SIGSOFT International Symposium on Software Testing and Analysis}}. \bibinfo{pages}{1022--1034}.
\newblock


\bibitem[{H}ugging {F}ace(2026)]%
        {ArceeaiLlama31}
\bibfield{author}{\bibinfo{person}{{H}ugging {F}ace}.} \bibinfo{year}{2026}\natexlab{}.
\newblock \bibinfo{title}{arcee-ai\/{L}lama-3.1-{S}uper{N}ova-{L}ite | {H}ugging {F}ace}.
\newblock \bibinfo{howpublished}{\url{https://huggingface.co/arcee-ai/{L}lama-3.1-{S}uper{N}ova-{L}ite}}.
\newblock
\newblock
\shownote{[Accessed 30-01-2026]}.


\bibitem[Humbatova et~al\mbox{.}(2020)]%
        {humbatova2020taxonomy}
\bibfield{author}{\bibinfo{person}{Nargiz Humbatova}, \bibinfo{person}{Gunel Jahangirova}, \bibinfo{person}{Gabriele Bavota}, \bibinfo{person}{Vincenzo Riccio}, \bibinfo{person}{Andrea Stocco}, {and} \bibinfo{person}{Paolo Tonella}.} \bibinfo{year}{2020}\natexlab{}.
\newblock \showarticletitle{Taxonomy of real faults in deep learning systems}. In \bibinfo{booktitle}{\emph{Proceedings of the ACM/IEEE 42nd international conference on software engineering}}. \bibinfo{pages}{1110--1121}.
\newblock


\bibitem[Inc(2026)]%
        {tldrlegal}
\bibfield{author}{\bibinfo{person}{FOSSA Inc}.} \bibinfo{year}{2026}\natexlab{}.
\newblock \bibinfo{title}{{T}{L}{D}{R}{L}egal | {S}oftware {L}icenses {E}xplained in {P}lain {E}nglish}.
\newblock \bibinfo{howpublished}{\url{https://www.tldrlegal.com}}.
\newblock
\newblock
\shownote{[Accessed 29-01-2026]}.


\bibitem[Intelligence(2026)]%
        {mordorintelligence}
\bibfield{author}{\bibinfo{person}{Mordor Intelligence}.} \bibinfo{year}{2026}\natexlab{}.
\newblock \bibinfo{title}{{L}arge {L}anguage {M}odel {M}arket {S}ize, {G}rowth \& {O}utlook | {I}ndustry {R}eport 2031}.
\newblock \bibinfo{howpublished}{\url{https://www.mordorintelligence.com/industry-reports/large-language-model-llm-market}}.
\newblock
\newblock
\shownote{[Accessed 28-01-2026]}.


\bibitem[Jahanshahi et~al\mbox{.}(2025)]%
        {11025699}
\bibfield{author}{\bibinfo{person}{Mahmoud Jahanshahi}, \bibinfo{person}{David Reid}, \bibinfo{person}{Adam McDaniel}, {and} \bibinfo{person}{Audris Mockus}.} \bibinfo{year}{2025}\natexlab{}.
\newblock \showarticletitle{OSS License Identification at Scale: A Comprehensive Dataset Using World of Code}. In \bibinfo{booktitle}{\emph{2025 IEEE/ACM 22nd International Conference on Mining Software Repositories (MSR)}}. \bibinfo{pages}{144--148}.
\newblock
\href{https://doi.org/10.1109/MSR66628.2025.00032}{doi:\nolinkurl{10.1109/MSR66628.2025.00032}}


\bibitem[Jalali and Wohlin(2012)]%
        {jalali2012systematic}
\bibfield{author}{\bibinfo{person}{Samireh Jalali} {and} \bibinfo{person}{Claes Wohlin}.} \bibinfo{year}{2012}\natexlab{}.
\newblock \showarticletitle{Systematic literature studies: database searches vs. backward snowballing}. In \bibinfo{booktitle}{\emph{Proceedings of the ACM-IEEE international symposium on Empirical software engineering and measurement}}. \bibinfo{pages}{29--38}.
\newblock


\bibitem[Jiang et~al\mbox{.}(2023a)]%
        {jiang2023empirical}
\bibfield{author}{\bibinfo{person}{Wenxin Jiang}, \bibinfo{person}{Nicholas Synovic}, \bibinfo{person}{Matt Hyatt}, \bibinfo{person}{Taylor~R Schorlemmer}, \bibinfo{person}{Rohan Sethi}, \bibinfo{person}{Yung-Hsiang Lu}, \bibinfo{person}{George~K Thiruvathukal}, {and} \bibinfo{person}{James~C Davis}.} \bibinfo{year}{2023}\natexlab{a}.
\newblock \showarticletitle{An empirical study of pre-trained model reuse in the hugging face deep learning model registry}. In \bibinfo{booktitle}{\emph{2023 IEEE/ACM 45th International Conference on Software Engineering (ICSE)}}. IEEE, \bibinfo{pages}{2463--2475}.
\newblock


\bibitem[Jiang et~al\mbox{.}(2023b)]%
        {10173952}
\bibfield{author}{\bibinfo{person}{Wenxin Jiang}, \bibinfo{person}{Nicholas Synovic}, \bibinfo{person}{Purvish Jajal}, \bibinfo{person}{Taylor~R. Schorlemmer}, \bibinfo{person}{Arav Tewari}, \bibinfo{person}{Bhavesh Pareek}, \bibinfo{person}{George~K. Thiruvathukal}, {and} \bibinfo{person}{James~C. Davis}.} \bibinfo{year}{2023}\natexlab{b}.
\newblock \showarticletitle{PTMTorrent: A Dataset for Mining Open-source Pre-trained Model Packages}. In \bibinfo{booktitle}{\emph{2023 IEEE/ACM 20th International Conference on Mining Software Repositories (MSR)}}. \bibinfo{pages}{57--61}.
\newblock
\href{https://doi.org/10.1109/MSR59073.2023.00021}{doi:\nolinkurl{10.1109/MSR59073.2023.00021}}


\bibitem[Jiang et~al\mbox{.}(2024)]%
        {jiang2024peatmoss}
\bibfield{author}{\bibinfo{person}{Wenxin Jiang}, \bibinfo{person}{Jerin Yasmin}, \bibinfo{person}{Jason Jones}, \bibinfo{person}{Nicholas Synovic}, \bibinfo{person}{Jiashen Kuo}, \bibinfo{person}{Nathaniel Bielanski}, \bibinfo{person}{Yuan Tian}, \bibinfo{person}{George~K Thiruvathukal}, {and} \bibinfo{person}{James~C Davis}.} \bibinfo{year}{2024}\natexlab{}.
\newblock \showarticletitle{Peatmoss: A dataset and initial analysis of pre-trained models in open-source software}. In \bibinfo{booktitle}{\emph{Proceedings of the 21st International Conference on Mining Software Repositories}}. \bibinfo{pages}{431--443}.
\newblock


\bibitem[Kapitsaki et~al\mbox{.}(2017)]%
        {kapitsaki2017automating}
\bibfield{author}{\bibinfo{person}{Georgia~M Kapitsaki}, \bibinfo{person}{Frederik Kramer}, {and} \bibinfo{person}{Nikolaos~D Tselikas}.} \bibinfo{year}{2017}\natexlab{}.
\newblock \showarticletitle{Automating the license compatibility process in open source software with SPDX}.
\newblock \bibinfo{journal}{\emph{Journal of systems and software}}  \bibinfo{volume}{131} (\bibinfo{year}{2017}), \bibinfo{pages}{386--401}.
\newblock


\bibitem[Kapitsaki and Paschalides(2017)]%
        {kapitsaki2017identifying}
\bibfield{author}{\bibinfo{person}{Georgia~M Kapitsaki} {and} \bibinfo{person}{Demetris Paschalides}.} \bibinfo{year}{2017}\natexlab{}.
\newblock \showarticletitle{Identifying terms in open source software license texts}. In \bibinfo{booktitle}{\emph{2017 24th Asia-Pacific Software Engineering Conference (APSEC)}}. IEEE, \bibinfo{pages}{540--545}.
\newblock


\bibitem[Karvelis et~al\mbox{.}(2018)]%
        {karvelis2018topic}
\bibfield{author}{\bibinfo{person}{Petros Karvelis}, \bibinfo{person}{Dimitris Gavrilis}, \bibinfo{person}{George Georgoulas}, {and} \bibinfo{person}{Chrysostomos Stylios}.} \bibinfo{year}{2018}\natexlab{}.
\newblock \showarticletitle{Topic recommendation using Doc2Vec}. In \bibinfo{booktitle}{\emph{2018 International Joint Conference on Neural Networks (IJCNN)}}. IEEE, \bibinfo{pages}{1--6}.
\newblock


\bibitem[Ke et~al\mbox{.}(2025)]%
        {ke2025clausebench}
\bibfield{author}{\bibinfo{person}{Qiang Ke}, \bibinfo{person}{Xinyi Hou}, \bibinfo{person}{Yanjie Zhao}, {and} \bibinfo{person}{Haoyu Wang}.} \bibinfo{year}{2025}\natexlab{}.
\newblock \showarticletitle{ClauseBench: Enhancing Software License Analysis with Clause-Level Benchmarking}. In \bibinfo{booktitle}{\emph{2025 IEEE/ACM 47th International Conference on Software Engineering: Companion Proceedings (ICSE-Companion)}}. IEEE, \bibinfo{pages}{255--266}.
\newblock


\bibitem[Krueger(1992)]%
        {10.1145/130844.130856}
\bibfield{author}{\bibinfo{person}{Charles~W. Krueger}.} \bibinfo{year}{1992}\natexlab{}.
\newblock \showarticletitle{Software reuse}.
\newblock \bibinfo{journal}{\emph{ACM Comput. Surv.}} \bibinfo{volume}{24}, \bibinfo{number}{2} (\bibinfo{date}{June} \bibinfo{year}{1992}), \bibinfo{pages}{131–183}.
\newblock
\showISSN{0360-0300}
\href{https://doi.org/10.1145/130844.130856}{doi:\nolinkurl{10.1145/130844.130856}}


\bibitem[Li et~al\mbox{.}(2025)]%
        {li2025open}
\bibfield{author}{\bibinfo{person}{Boyuan Li}, \bibinfo{person}{Chengwei Liu}, \bibinfo{person}{Lingling Fan}, \bibinfo{person}{Sen Chen}, \bibinfo{person}{Zhenlin Zhang}, {and} \bibinfo{person}{Zheli Liu}.} \bibinfo{year}{2025}\natexlab{}.
\newblock \showarticletitle{Open Source, Hidden Costs: A Systematic Literature Review on OSS License Management}.
\newblock \bibinfo{journal}{\emph{IEEE Transactions on Software Engineering}} (\bibinfo{year}{2025}).
\newblock


\bibitem[Li et~al\mbox{.}(2022)]%
        {li2022scalpel}
\bibfield{author}{\bibinfo{person}{Li Li}, \bibinfo{person}{Jiawei Wang}, {and} \bibinfo{person}{Haowei Quan}.} \bibinfo{year}{2022}\natexlab{}.
\newblock \showarticletitle{Scalpel: The Python Static Analysis Framework}.
\newblock \bibinfo{journal}{\emph{arXiv preprint arXiv:2202.11840}} (\bibinfo{year}{2022}).
\newblock


\bibitem[Li et~al\mbox{.}(2023)]%
        {li2023lisum}
\bibfield{author}{\bibinfo{person}{Linyu Li}, \bibinfo{person}{Sihan Xu}, \bibinfo{person}{Yang Liu}, \bibinfo{person}{Ya Gao}, \bibinfo{person}{Xiangrui Cai}, \bibinfo{person}{Jiarun Wu}, \bibinfo{person}{Wenli Song}, {and} \bibinfo{person}{Zheli Liu}.} \bibinfo{year}{2023}\natexlab{}.
\newblock \showarticletitle{LiSum: Open Source Software License Summarization with Multi-Task Learning}. In \bibinfo{booktitle}{\emph{2023 38th IEEE/ACM International Conference on Automated Software Engineering (ASE)}}. IEEE, \bibinfo{pages}{787--799}.
\newblock


\bibitem[Liu et~al\mbox{.}(2024a)]%
        {liu2024deepseek}
\bibfield{author}{\bibinfo{person}{Aixin Liu}, \bibinfo{person}{Bei Feng}, \bibinfo{person}{Bing Xue}, \bibinfo{person}{Bingxuan Wang}, \bibinfo{person}{Bochao Wu}, \bibinfo{person}{Chengda Lu}, \bibinfo{person}{Chenggang Zhao}, \bibinfo{person}{Chengqi Deng}, \bibinfo{person}{Chenyu Zhang}, \bibinfo{person}{Chong Ruan}, {et~al\mbox{.}}} \bibinfo{year}{2024}\natexlab{a}.
\newblock \showarticletitle{Deepseek-v3 technical report}.
\newblock \bibinfo{journal}{\emph{arXiv preprint arXiv:2412.19437}} (\bibinfo{year}{2024}).
\newblock


\bibitem[Liu et~al\mbox{.}(2023)]%
        {liu2023towards}
\bibfield{author}{\bibinfo{person}{Di Liu}, \bibinfo{person}{Yang Feng}, \bibinfo{person}{Yanyan Yan}, {and} \bibinfo{person}{Baowen Xu}.} \bibinfo{year}{2023}\natexlab{}.
\newblock \showarticletitle{Towards understanding bugs in python interpreters}.
\newblock \bibinfo{journal}{\emph{Empirical Software Engineering}} \bibinfo{volume}{28}, \bibinfo{number}{1} (\bibinfo{year}{2023}), \bibinfo{pages}{19}.
\newblock


\bibitem[Liu et~al\mbox{.}(2024c)]%
        {liu2024large}
\bibfield{author}{\bibinfo{person}{Junwei Liu}, \bibinfo{person}{Kaixin Wang}, \bibinfo{person}{Yixuan Chen}, \bibinfo{person}{Xin Peng}, \bibinfo{person}{Zhenpeng Chen}, \bibinfo{person}{Lingming Zhang}, {and} \bibinfo{person}{Yiling Lou}.} \bibinfo{year}{2024}\natexlab{c}.
\newblock \showarticletitle{Large language model-based agents for software engineering: A survey}.
\newblock \bibinfo{journal}{\emph{arXiv preprint arXiv:2409.02977}} (\bibinfo{year}{2024}).
\newblock


\bibitem[Liu et~al\mbox{.}(2025)]%
        {liu2025empiricalstudyvulnerablepackage}
\bibfield{author}{\bibinfo{person}{Shuhan Liu}, \bibinfo{person}{Xing Hu}, \bibinfo{person}{Xin Xia}, \bibinfo{person}{David Lo}, {and} \bibinfo{person}{Xiaohu Yang}.} \bibinfo{year}{2025}\natexlab{}.
\newblock \bibinfo{title}{An Empirical Study of Vulnerable Package Dependencies in LLM Repositories}.
\newblock
\showeprint[arxiv]{2508.21417}~[cs.CR]
\urldef\tempurl%
\url{https://arxiv.org/abs/2508.21417}
\showURL{%
\tempurl}


\bibitem[Liu et~al\mbox{.}(2024b)]%
        {liu2024catch}
\bibfield{author}{\bibinfo{person}{Tao Liu}, \bibinfo{person}{Chengwei Liu}, \bibinfo{person}{Tianwei Liu}, \bibinfo{person}{He Wang}, \bibinfo{person}{Gaofei Wu}, \bibinfo{person}{Yang Liu}, {and} \bibinfo{person}{Yuqing Zhang}.} \bibinfo{year}{2024}\natexlab{b}.
\newblock \showarticletitle{Catch the butterfly: Peeking into the terms and conflicts among spdx licenses}. In \bibinfo{booktitle}{\emph{2024 IEEE International Conference on Software Analysis, Evolution and Reengineering (SANER)}}. IEEE, \bibinfo{pages}{477--488}.
\newblock


\bibitem[Liu et~al\mbox{.}(2019)]%
        {liu2019predicting}
\bibfield{author}{\bibinfo{person}{Xiaoyu Liu}, \bibinfo{person}{LiGuo Huang}, \bibinfo{person}{Jidong Ge}, {and} \bibinfo{person}{Vincent Ng}.} \bibinfo{year}{2019}\natexlab{}.
\newblock \showarticletitle{Predicting licenses for changed source code}. In \bibinfo{booktitle}{\emph{2019 34th IEEE/ACM International Conference on Automated Software Engineering (ASE)}}. IEEE, \bibinfo{pages}{686--697}.
\newblock


\bibitem[Liu et~al\mbox{.}(2026)]%
        {liu2026towards}
\bibfield{author}{\bibinfo{person}{Ziang Liu}, \bibinfo{person}{Xin Liu}, \bibinfo{person}{Yingli Zhang}, \bibinfo{person}{Song Li}, \bibinfo{person}{Weina Niu}, \bibinfo{person}{Qingguo Zhou}, \bibinfo{person}{Rui Zhou}, {and} \bibinfo{person}{Xiaokang Zhou}.} \bibinfo{year}{2026}\natexlab{}.
\newblock \showarticletitle{Towards a comprehensive framework for verifying open-source software license compatibility}.
\newblock \bibinfo{journal}{\emph{Empirical Software Engineering}} \bibinfo{volume}{31}, \bibinfo{number}{1} (\bibinfo{year}{2026}), \bibinfo{pages}{15}.
\newblock


\bibitem[Macklon et~al\mbox{.}(2023)]%
        {macklon2023taxonomy}
\bibfield{author}{\bibinfo{person}{Finlay Macklon}, \bibinfo{person}{Markos Viggiato}, \bibinfo{person}{Natalia Romanova}, \bibinfo{person}{Chris Buzon}, \bibinfo{person}{Dale Paas}, {and} \bibinfo{person}{Cor-Paul Bezemer}.} \bibinfo{year}{2023}\natexlab{}.
\newblock \showarticletitle{A Taxonomy of Testable HTML5 Canvas Issues}.
\newblock \bibinfo{journal}{\emph{IEEE Transactions on Software Engineering}} \bibinfo{volume}{49}, \bibinfo{number}{6} (\bibinfo{year}{2023}), \bibinfo{pages}{3647--3659}.
\newblock


\bibitem[Mancinelli et~al\mbox{.}(2006)]%
        {mancinelli2006managing}
\bibfield{author}{\bibinfo{person}{Fabio Mancinelli}, \bibinfo{person}{Jaap Boender}, \bibinfo{person}{Roberto Di~Cosmo}, \bibinfo{person}{Jerome Vouillon}, \bibinfo{person}{Berke Durak}, \bibinfo{person}{Xavier Leroy}, {and} \bibinfo{person}{Ralf Treinen}.} \bibinfo{year}{2006}\natexlab{}.
\newblock \showarticletitle{Managing the complexity of large free and open source package-based software distributions}. In \bibinfo{booktitle}{\emph{21st IEEE/ACM International Conference on Automated Software Engineering (ASE'06)}}. IEEE, \bibinfo{pages}{199--208}.
\newblock


\bibitem[Mathur et~al\mbox{.}(2012)]%
        {mathur2012empirical}
\bibfield{author}{\bibinfo{person}{Arunesh Mathur}, \bibinfo{person}{Harshal Choudhary}, \bibinfo{person}{Priyank Vashist}, \bibinfo{person}{William Thies}, {and} \bibinfo{person}{Santhi Thilagam}.} \bibinfo{year}{2012}\natexlab{}.
\newblock \showarticletitle{An empirical study of license violations in open source projects}. In \bibinfo{booktitle}{\emph{2012 35th annual IEEE software engineering workshop}}. IEEE, \bibinfo{pages}{168--176}.
\newblock


\bibitem[Munaiah et~al\mbox{.}(2017)]%
        {DBLP:journals/ese/MunaiahKCN17}
\bibfield{author}{\bibinfo{person}{Nuthan Munaiah}, \bibinfo{person}{Steven Kroh}, \bibinfo{person}{Craig Cabrey}, {and} \bibinfo{person}{Meiyappan Nagappan}.} \bibinfo{year}{2017}\natexlab{}.
\newblock \showarticletitle{Curating GitHub for engineered software projects}.
\newblock \bibinfo{journal}{\emph{Empir. Softw. Eng.}} \bibinfo{volume}{22}, \bibinfo{number}{6} (\bibinfo{year}{2017}), \bibinfo{pages}{3219--3253}.
\newblock
\href{https://doi.org/10.1007/S10664-017-9512-6}{doi:\nolinkurl{10.1007/S10664-017-9512-6}}


\bibitem[Niu et~al\mbox{.}(2026)]%
        {niu2026trustinsecuredemystifyingdevelopers}
\bibfield{author}{\bibinfo{person}{Yuqing Niu}, \bibinfo{person}{Jieke Shi}, \bibinfo{person}{Ruidong Han}, \bibinfo{person}{Ye Liu}, \bibinfo{person}{Chengyan Ma}, \bibinfo{person}{Yunbo Lyu}, {and} \bibinfo{person}{David Lo}.} \bibinfo{year}{2026}\natexlab{}.
\newblock \bibinfo{title}{What You Trust Is Insecure: Demystifying How Developers (Mis)Use Trusted Execution Environments in Practice}.
\newblock
\showeprint[arxiv]{2512.17363}~[cs.SE]
\urldef\tempurl%
\url{https://arxiv.org/abs/2512.17363}
\showURL{%
\tempurl}


\bibitem[OpenAI(2026)]%
        {chatgpt}
\bibfield{author}{\bibinfo{person}{OpenAI}.} \bibinfo{year}{2026}\natexlab{}.
\newblock \bibinfo{title}{{C}hat{G}{P}{T}}.
\newblock \bibinfo{howpublished}{\url{https://chatgpt.com}}.
\newblock
\newblock
\shownote{[Accessed 28-01-2026]}.


\bibitem[Papoutsoglou et~al\mbox{.}(2022)]%
        {papoutsoglou2022analysis}
\bibfield{author}{\bibinfo{person}{Maria Papoutsoglou}, \bibinfo{person}{Georgia~M Kapitsaki}, \bibinfo{person}{Daniel German}, {and} \bibinfo{person}{Lefteris Angelis}.} \bibinfo{year}{2022}\natexlab{}.
\newblock \showarticletitle{An analysis of open source software licensing questions in stack exchange sites}.
\newblock \bibinfo{journal}{\emph{Journal of Systems and Software}}  \bibinfo{volume}{183} (\bibinfo{year}{2022}), \bibinfo{pages}{111113}.
\newblock


\bibitem[Shi et~al\mbox{.}(2025)]%
        {10.1145/3708525}
\bibfield{author}{\bibinfo{person}{Jieke Shi}, \bibinfo{person}{Zhou Yang}, {and} \bibinfo{person}{David Lo}.} \bibinfo{year}{2025}\natexlab{}.
\newblock \showarticletitle{Efficient and Green Large Language Models for Software Engineering: Literature Review, Vision, and the Road Ahead}.
\newblock \bibinfo{journal}{\emph{ACM Trans. Softw. Eng. Methodol.}} \bibinfo{volume}{34}, \bibinfo{number}{5}, Article \bibinfo{articleno}{137} (\bibinfo{date}{May} \bibinfo{year}{2025}), \bibinfo{numpages}{22}~pages.
\newblock
\showISSN{1049-331X}
\href{https://doi.org/10.1145/3708525}{doi:\nolinkurl{10.1145/3708525}}


\bibitem[Socher et~al\mbox{.}(2013)]%
        {socher2013recursive}
\bibfield{author}{\bibinfo{person}{Richard Socher}, \bibinfo{person}{Alex Perelygin}, \bibinfo{person}{Jean Wu}, \bibinfo{person}{Jason Chuang}, \bibinfo{person}{Christopher~D Manning}, \bibinfo{person}{Andrew~Y Ng}, {and} \bibinfo{person}{Christopher Potts}.} \bibinfo{year}{2013}\natexlab{}.
\newblock \showarticletitle{Recursive deep models for semantic compositionality over a sentiment treebank}. In \bibinfo{booktitle}{\emph{Proceedings of the 2013 conference on empirical methods in natural language processing}}. \bibinfo{pages}{1631--1642}.
\newblock


\bibitem[{S}ourcegraph(2026)]%
        {sourcegraph}
\bibfield{author}{\bibinfo{person}{{S}ourcegraph}.} \bibinfo{year}{2026}\natexlab{}.
\newblock \bibinfo{title}{{S}ourcegraph | {T}he code intelligence platform for enterprises --- sourcegraph.com}.
\newblock \bibinfo{howpublished}{\url{https://sourcegraph.com}}.
\newblock
\newblock
\shownote{[Accessed 29-01-2026]}.


\bibitem[SPDX(2026)]%
        {spdx}
\bibfield{author}{\bibinfo{person}{SPDX}.} \bibinfo{year}{2026}\natexlab{}.
\newblock \bibinfo{title}{{S}{P}{D}{X} {L}icense {L}ist | {S}oftware {P}ackage {D}ata {E}xchange ({S}{P}{D}{X})}.
\newblock \bibinfo{howpublished}{\url{https://spdx.org/licenses/}}.
\newblock
\newblock
\shownote{[Accessed 29-01-2026]}.


\bibitem[Stewart et~al\mbox{.}(2010)]%
        {stewart2010software}
\bibfield{author}{\bibinfo{person}{Kate Stewart}, \bibinfo{person}{Phil Odence}, {and} \bibinfo{person}{Esteban Rockett}.} \bibinfo{year}{2010}\natexlab{}.
\newblock \showarticletitle{Software package data exchange (SPDX) specification}.
\newblock \bibinfo{journal}{\emph{IFOSS L. Rev.}}  \bibinfo{volume}{2} (\bibinfo{year}{2010}), \bibinfo{pages}{191}.
\newblock


\bibitem[Tan et~al\mbox{.}(2014)]%
        {tan2014bug}
\bibfield{author}{\bibinfo{person}{Lin Tan}, \bibinfo{person}{Chen Liu}, \bibinfo{person}{Zhenmin Li}, \bibinfo{person}{Xuanhui Wang}, \bibinfo{person}{Yuanyuan Zhou}, {and} \bibinfo{person}{Chengxiang Zhai}.} \bibinfo{year}{2014}\natexlab{}.
\newblock \showarticletitle{Bug characteristics in open source software}.
\newblock \bibinfo{journal}{\emph{Empirical software engineering}}  \bibinfo{volume}{19} (\bibinfo{year}{2014}), \bibinfo{pages}{1665--1705}.
\newblock


\bibitem[Terragni et~al\mbox{.}(2025)]%
        {terragni2025future}
\bibfield{author}{\bibinfo{person}{Valerio Terragni}, \bibinfo{person}{Annie Vella}, \bibinfo{person}{Partha Roop}, {and} \bibinfo{person}{Kelly Blincoe}.} \bibinfo{year}{2025}\natexlab{}.
\newblock \showarticletitle{The future of AI-driven software engineering}.
\newblock \bibinfo{journal}{\emph{ACM Transactions on Software Engineering and Methodology}} \bibinfo{volume}{34}, \bibinfo{number}{5} (\bibinfo{year}{2025}), \bibinfo{pages}{1--20}.
\newblock


\bibitem[Vaswani et~al\mbox{.}(2017)]%
        {vaswani2017attention}
\bibfield{author}{\bibinfo{person}{Ashish Vaswani}, \bibinfo{person}{Noam Shazeer}, \bibinfo{person}{Niki Parmar}, \bibinfo{person}{Jakob Uszkoreit}, \bibinfo{person}{Llion Jones}, \bibinfo{person}{Aidan~N Gomez}, \bibinfo{person}{{\L}ukasz Kaiser}, {and} \bibinfo{person}{Illia Polosukhin}.} \bibinfo{year}{2017}\natexlab{}.
\newblock \showarticletitle{Attention is all you need}.
\newblock \bibinfo{journal}{\emph{Advances in neural information processing systems}}  \bibinfo{volume}{30} (\bibinfo{year}{2017}).
\newblock


\bibitem[Viera et~al\mbox{.}(2005)]%
        {viera2005understanding}
\bibfield{author}{\bibinfo{person}{Anthony~J Viera}, \bibinfo{person}{Joanne~M Garrett}, {et~al\mbox{.}}} \bibinfo{year}{2005}\natexlab{}.
\newblock \showarticletitle{Understanding interobserver agreement: the kappa statistic}.
\newblock \bibinfo{journal}{\emph{Fam med}} \bibinfo{volume}{37}, \bibinfo{number}{5} (\bibinfo{year}{2005}), \bibinfo{pages}{360--363}.
\newblock


\bibitem[Wang et~al\mbox{.}(2025a)]%
        {wang2025comprehensive}
\bibfield{author}{\bibinfo{person}{Bo Wang}, \bibinfo{person}{Chong Chen}, \bibinfo{person}{Junjie Chen}, \bibinfo{person}{Bowen Xu}, \bibinfo{person}{Chen Ye}, \bibinfo{person}{Youfang Lin}, \bibinfo{person}{Guoliang Dong}, {and} \bibinfo{person}{Jun Sun}.} \bibinfo{year}{2025}\natexlab{a}.
\newblock \showarticletitle{A Comprehensive Study of OOP-Related Bugs in C++ Compilers}.
\newblock \bibinfo{journal}{\emph{IEEE Transactions on Software Engineering}} (\bibinfo{year}{2025}).
\newblock


\bibitem[Wang et~al\mbox{.}(2024)]%
        {wang2024software}
\bibfield{author}{\bibinfo{person}{Junjie Wang}, \bibinfo{person}{Yuchao Huang}, \bibinfo{person}{Chunyang Chen}, \bibinfo{person}{Zhe Liu}, \bibinfo{person}{Song Wang}, {and} \bibinfo{person}{Qing Wang}.} \bibinfo{year}{2024}\natexlab{}.
\newblock \showarticletitle{Software testing with large language models: Survey, landscape, and vision}.
\newblock \bibinfo{journal}{\emph{IEEE Transactions on Software Engineering}} (\bibinfo{year}{2024}).
\newblock


\bibitem[Wang et~al\mbox{.}(2025b)]%
        {10.1145/3708531}
\bibfield{author}{\bibinfo{person}{Shenao Wang}, \bibinfo{person}{Yanjie Zhao}, \bibinfo{person}{Xinyi Hou}, {and} \bibinfo{person}{Haoyu Wang}.} \bibinfo{year}{2025}\natexlab{b}.
\newblock \showarticletitle{Large Language Model Supply Chain: A Research Agenda}.
\newblock \bibinfo{journal}{\emph{ACM Trans. Softw. Eng. Methodol.}} \bibinfo{volume}{34}, \bibinfo{number}{5}, Article \bibinfo{articleno}{147} (\bibinfo{date}{May} \bibinfo{year}{2025}), \bibinfo{numpages}{46}~pages.
\newblock
\showISSN{1049-331X}
\href{https://doi.org/10.1145/3708531}{doi:\nolinkurl{10.1145/3708531}}


\bibitem[Williams et~al\mbox{.}(2025)]%
        {10.1145/3714464}
\bibfield{author}{\bibinfo{person}{Laurie Williams}, \bibinfo{person}{Giacomo Benedetti}, \bibinfo{person}{Sivana Hamer}, \bibinfo{person}{Ranindya Paramitha}, \bibinfo{person}{Imranur Rahman}, \bibinfo{person}{Mahzabin Tamanna}, \bibinfo{person}{Greg Tystahl}, \bibinfo{person}{Nusrat Zahan}, \bibinfo{person}{Patrick Morrison}, \bibinfo{person}{Yasemin Acar}, \bibinfo{person}{Michel Cukier}, \bibinfo{person}{Christian K\"{a}stner}, \bibinfo{person}{Alexandros Kapravelos}, \bibinfo{person}{Dominik Wermke}, {and} \bibinfo{person}{William Enck}.} \bibinfo{year}{2025}\natexlab{}.
\newblock \showarticletitle{Research Directions in Software Supply Chain Security}.
\newblock \bibinfo{journal}{\emph{ACM Trans. Softw. Eng. Methodol.}} \bibinfo{volume}{34}, \bibinfo{number}{5}, Article \bibinfo{articleno}{146} (\bibinfo{date}{May} \bibinfo{year}{2025}), \bibinfo{numpages}{38}~pages.
\newblock
\showISSN{1049-331X}
\href{https://doi.org/10.1145/3714464}{doi:\nolinkurl{10.1145/3714464}}


\bibitem[Wohlin(2014)]%
        {wohlin2014guidelines}
\bibfield{author}{\bibinfo{person}{Claes Wohlin}.} \bibinfo{year}{2014}\natexlab{}.
\newblock \showarticletitle{Guidelines for snowballing in systematic literature studies and a replication in software engineering}. In \bibinfo{booktitle}{\emph{Proceedings of the 18th international conference on evaluation and assessment in software engineering}}. \bibinfo{pages}{1--10}.
\newblock


\bibitem[Wu et~al\mbox{.}(2024)]%
        {wu2024large}
\bibfield{author}{\bibinfo{person}{Jiaqi Wu}, \bibinfo{person}{Lingfeng Bao}, \bibinfo{person}{Xiaohu Yang}, \bibinfo{person}{Xin Xia}, {and} \bibinfo{person}{Xing Hu}.} \bibinfo{year}{2024}\natexlab{}.
\newblock \showarticletitle{A large-scale empirical study of open source license usage: Practices and challenges}. In \bibinfo{booktitle}{\emph{Proceedings of the 21st International Conference on Mining Software Repositories}}. \bibinfo{pages}{595--606}.
\newblock


\bibitem[Xu et~al\mbox{.}(2023a)]%
        {xu2023liresolver}
\bibfield{author}{\bibinfo{person}{Sihan Xu}, \bibinfo{person}{Ya Gao}, \bibinfo{person}{Lingling Fan}, \bibinfo{person}{Linyu Li}, \bibinfo{person}{Xiangrui Cai}, {and} \bibinfo{person}{Zheli Liu}.} \bibinfo{year}{2023}\natexlab{a}.
\newblock \showarticletitle{Liresolver: License incompatibility resolution for open source software}. In \bibinfo{booktitle}{\emph{Proceedings of the 32nd ACM SIGSOFT International Symposium on Software Testing and Analysis}}. \bibinfo{pages}{652--663}.
\newblock


\bibitem[Xu et~al\mbox{.}(2023b)]%
        {xu2023lidetector}
\bibfield{author}{\bibinfo{person}{Sihan Xu}, \bibinfo{person}{Ya Gao}, \bibinfo{person}{Lingling Fan}, \bibinfo{person}{Zheli Liu}, \bibinfo{person}{Yang Liu}, {and} \bibinfo{person}{Hua Ji}.} \bibinfo{year}{2023}\natexlab{b}.
\newblock \showarticletitle{Lidetector: License incompatibility detection for open source software}.
\newblock \bibinfo{journal}{\emph{ACM Transactions on Software Engineering and Methodology}} \bibinfo{volume}{32}, \bibinfo{number}{1} (\bibinfo{year}{2023}), \bibinfo{pages}{1--28}.
\newblock


\bibitem[Xu et~al\mbox{.}(2024a)]%
        {xu2024licoeval}
\bibfield{author}{\bibinfo{person}{Weiwei Xu}, \bibinfo{person}{Kai Gao}, \bibinfo{person}{Hao He}, {and} \bibinfo{person}{Minghui Zhou}.} \bibinfo{year}{2024}\natexlab{a}.
\newblock \showarticletitle{Licoeval: Evaluating llms on license compliance in code generation}.
\newblock \bibinfo{journal}{\emph{arXiv preprint arXiv:2408.02487}} (\bibinfo{year}{2024}).
\newblock


\bibitem[Xu et~al\mbox{.}(2025)]%
        {xu2025first}
\bibfield{author}{\bibinfo{person}{Weiwei Xu}, \bibinfo{person}{Hengzhi Ye}, \bibinfo{person}{Kai Gao}, {and} \bibinfo{person}{Minghui Zhou}.} \bibinfo{year}{2025}\natexlab{}.
\newblock \showarticletitle{A first look at License Variants in the PyPI Ecosystem}.
\newblock \bibinfo{journal}{\emph{arXiv preprint arXiv:2507.14594}} (\bibinfo{year}{2025}).
\newblock


\bibitem[Xu et~al\mbox{.}(2024b)]%
        {10629039}
\bibfield{author}{\bibinfo{person}{Zhenhua Xu}, \bibinfo{person}{Yujia Zhang}, \bibinfo{person}{Enze Xie}, \bibinfo{person}{Zhen Zhao}, \bibinfo{person}{Yong Guo}, \bibinfo{person}{Kwan-Yee~K. Wong}, \bibinfo{person}{Zhenguo Li}, {and} \bibinfo{person}{Hengshuang Zhao}.} \bibinfo{year}{2024}\natexlab{b}.
\newblock \showarticletitle{DriveGPT4: Interpretable End-to-End Autonomous Driving Via Large Language Model}.
\newblock \bibinfo{journal}{\emph{IEEE Robotics and Automation Letters}} \bibinfo{volume}{9}, \bibinfo{number}{10} (\bibinfo{year}{2024}), \bibinfo{pages}{8186--8193}.
\newblock
\href{https://doi.org/10.1109/LRA.2024.3440097}{doi:\nolinkurl{10.1109/LRA.2024.3440097}}


\bibitem[Yang et~al\mbox{.}(2025a)]%
        {yang2025qwen3}
\bibfield{author}{\bibinfo{person}{An Yang}, \bibinfo{person}{Anfeng Li}, \bibinfo{person}{Baosong Yang}, \bibinfo{person}{Beichen Zhang}, \bibinfo{person}{Binyuan Hui}, \bibinfo{person}{Bo Zheng}, \bibinfo{person}{Bowen Yu}, \bibinfo{person}{Chang Gao}, \bibinfo{person}{Chengen Huang}, \bibinfo{person}{Chenxu Lv}, {et~al\mbox{.}}} \bibinfo{year}{2025}\natexlab{a}.
\newblock \showarticletitle{Qwen3 technical report}.
\newblock \bibinfo{journal}{\emph{arXiv preprint arXiv:2505.09388}} (\bibinfo{year}{2025}).
\newblock


\bibitem[Yang et~al\mbox{.}(2025b)]%
        {yang2025ecosystem}
\bibfield{author}{\bibinfo{person}{Zhou Yang}, \bibinfo{person}{Jieke Shi}, \bibinfo{person}{Prem Devanbu}, {and} \bibinfo{person}{David Lo}.} \bibinfo{year}{2025}\natexlab{b}.
\newblock \showarticletitle{Ecosystem of Large Language Models for Code}.
\newblock \bibinfo{journal}{\emph{ACM Trans. Softw. Eng. Methodol.}} (\bibinfo{date}{April} \bibinfo{year}{2025}).
\newblock
\showISSN{1049-331X}
\href{https://doi.org/10.1145/3731753}{doi:\nolinkurl{10.1145/3731753}}
\newblock
\shownote{Just Accepted}.


\bibitem[Yi et~al\mbox{.}(2025)]%
        {10.1145/3771090}
\bibfield{author}{\bibinfo{person}{Zihao Yi}, \bibinfo{person}{Jiarui Ouyang}, \bibinfo{person}{Zhe Xu}, \bibinfo{person}{Yuwen Liu}, \bibinfo{person}{Tianhao Liao}, \bibinfo{person}{Haohao Luo}, {and} \bibinfo{person}{Ying Shen}.} \bibinfo{year}{2025}\natexlab{}.
\newblock \showarticletitle{A Survey on Recent Advances in LLM-Based Multi-turn Dialogue Systems}.
\newblock \bibinfo{journal}{\emph{ACM Comput. Surv.}} \bibinfo{volume}{58}, \bibinfo{number}{6}, Article \bibinfo{articleno}{148} (\bibinfo{date}{Dec.} \bibinfo{year}{2025}), \bibinfo{numpages}{38}~pages.
\newblock
\showISSN{0360-0300}
\href{https://doi.org/10.1145/3771090}{doi:\nolinkurl{10.1145/3771090}}


\bibitem[Zhang et~al\mbox{.}(2025)]%
        {10.1145/3696630.3728525}
\bibfield{author}{\bibinfo{person}{Lyuye Zhang}, \bibinfo{person}{Chengwei Liu}, \bibinfo{person}{Jiahui Wu}, \bibinfo{person}{Shiyang Zhang}, \bibinfo{person}{Chengyue Liu}, \bibinfo{person}{Zhengzi Xu}, \bibinfo{person}{Sen Chen}, {and} \bibinfo{person}{Yang Liu}.} \bibinfo{year}{2025}\natexlab{}.
\newblock \bibinfo{booktitle}{\emph{Drop the Golden Apples: Identifying Third-Party Reuse by DB-Less Software Composition Analysis}}.
\newblock \bibinfo{publisher}{Association for Computing Machinery}, \bibinfo{address}{New York, NY, USA}, \bibinfo{pages}{691–695}.
\newblock
\showISBNx{9798400712760}
\urldef\tempurl%
\url{https://doi.org/10.1145/3696630.3728525}
\showURL{%
\tempurl}


\bibitem[Zhao et~al\mbox{.}(2025)]%
        {zhao2025surveylargelanguagemodels}
\bibfield{author}{\bibinfo{person}{Wayne~Xin Zhao}, \bibinfo{person}{Kun Zhou}, \bibinfo{person}{Junyi Li}, \bibinfo{person}{Tianyi Tang}, \bibinfo{person}{Xiaolei Wang}, \bibinfo{person}{Yupeng Hou}, \bibinfo{person}{Yingqian Min}, \bibinfo{person}{Beichen Zhang}, \bibinfo{person}{Junjie Zhang}, \bibinfo{person}{Zican Dong}, \bibinfo{person}{Yifan Du}, \bibinfo{person}{Chen Yang}, \bibinfo{person}{Yushuo Chen}, \bibinfo{person}{Zhipeng Chen}, \bibinfo{person}{Jinhao Jiang}, \bibinfo{person}{Ruiyang Ren}, \bibinfo{person}{Yifan Li}, \bibinfo{person}{Xinyu Tang}, \bibinfo{person}{Zikang Liu}, \bibinfo{person}{Peiyu Liu}, \bibinfo{person}{Jian-Yun Nie}, {and} \bibinfo{person}{Ji-Rong Wen}.} \bibinfo{year}{2025}\natexlab{}.
\newblock \bibinfo{title}{A Survey of Large Language Models}.
\newblock
\showeprint[arxiv]{2303.18223}~[cs.CL]
\urldef\tempurl%
\url{https://arxiv.org/abs/2303.18223}
\showURL{%
\tempurl}


\end{thebibliography}

\end{document}